\documentclass[aps,pre,reprint,groupedaddress,showpacs]{revtex4-1} \usepackage{graphicx}
\usepackage{amsmath}

\usepackage{amssymb}
\usepackage{subfigure}

\usepackage{epsfig}
\usepackage{xcolor}
\usepackage{epstopdf}
\usepackage{bigints}
\usepackage{makecell}
\usepackage[title]{appendix}
\begin{document}
\title{Cooperative dynamics in bidirectional transport on flexible lattice}

\author{Akriti Jindal, Atul Kumar Verma}

\author{Arvind Kumar Gupta}
\email[]{akgupta@iitrpr.ac.in}

\affiliation{Department of Mathematics, Indian Institute of Technology Ropar, Rupnagar-140001, Punjab, India.}

\begin{abstract}
Several theoretical models based on totally asymmetric simple exclusion process (TASEP) have been extensively utilized to study various non-equilibrium transport phenomena. Inspired by the
the role of microtubule-transported vesicles in intracellular transport, we propose a generalized TASEP model where two distinct particles are directed to hop stochastically in opposite directions on a flexible lattice immersed in a three dimensional pool of diffusing particles. We investigate the interplay between lattice conformation and bidirectional transport by obtaining the stationary phase diagrams and density profiles within the framework of mean field theory. For the case when recycling strength is independent of density of particles, the topology of phase diagram alters quantitatively. However, if the lattice occupancy governs the global conformation of lattice, in addition to the pre-existing phases for bidirectional transport a new asymmetric shock-low density phase originates in the system. We identified that this phase is sensitive to finite size effect and vanishes in the thermodynamic limit.
\end{abstract}


\maketitle

\section{Introduction}\label{sec1}
Over the decade, stochastic transport processes have been studied intensively relying on the variations of totally asymmetric simple exclusion process (TASEP) \cite{widom1991repton,chowdhury2000statistical,belitsky2001cellular,foulaadvand2016phase}. Despite its simplicity, the development of exclusion processes led to the deeper understanding of complex non-equilibrium phenomenon. Amongst others, several biological transport processes have been examined where it was originally proposed in 1968 to study the kinetics of biopolymerization \cite{macdonald1968kinetics}. Since, then it has also been a discipline of significant interest to convey the dynamics of motor proteins progressing on a microtubule.
In TASEP, the active species are considered as point particles that are allowed to enter and exit the lattice from the two extreme ends and hop along a preferred direction with some rate obeying hard core exclusion principle. A single species model has been extensively studied in the literature \cite{blythe2007nonequilibrium,zia2011modeling} that explained many complex phenomenon such as boundary-induced phase transitions, phase separation,
shock formation \cite{kolomeisky1998phase,evans2003shock}.
\begin{figure*}[!htb]
        \center{\includegraphics[trim = 150 480 120 70,width=8cm]{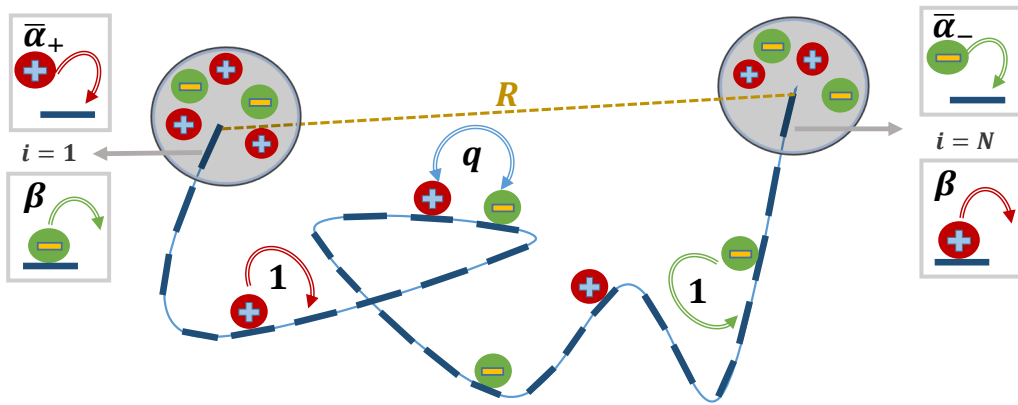}}
        \caption{\label{fig1} Sketch of TASEP with two different type of particles denoted by `$+$' and `$-$' moving along a flexible lattice of length $i=1,2,\hdots,N$. `$+$', `$-$' particles enter the lattice with effective entry rate $\bar{\alpha}_{+}$ and $\bar{\alpha}_{-}$ respectively and leave the lattice with rate $\beta$. Particles hop with unit rate in the bulk and swap their positions with rate $q$ when the two consecutive sites are simultaneously occupied by two distinct particles. $R$ represents end-to-end distance between two extreme ends of a lattice.}
\end{figure*}

In intracellular transport, the microtubules (MTs) are known to serve as an important component that are laid down for the mobility of motor proteins contributing to the majority of functions in eukrayotic cells \cite{soppina2009tug}. The structure of motors and polarity of MTs direct the movement of motor proteins. The oppositely directed motor proteins e.g. kinesin and dynein move along a filament to carry cellular cargo such as lipid droplets, mitochondria, endosomes from one place to another in the cell \cite{hancock2014bidirectional}. The majority of kinesin motors proceed towards the plus end of MT whereas dynein motors are directed to minus end of MT \cite{hancock2014bidirectional,soppina2009tug}. The functioning of motor proteins is essential for the survival of a cell because its mutation and disruption can lead to development of diseases such as Alzheimer's, neurodegenrative and polycicstic kidney diseases \cite{schliwa2003molecular}. Generalizing a single species model, previous theoretical attempts focused to study multiple species model that revealed a variety of cooperative phenomenon such as symmetry breaking and phase separation \cite{evans1995spontaneous,sharma2017phase,verma2018far}. The ``bridge model" was the first model to exhibit spontaneous symmetry breaking (SSB) where phases with broken symmetry were reported under symmetrical conditions of two species \cite{evans1995spontaneous,evans1995asymmetric}.

So far, for the theoretical studies, MTs have been considered to be linear rigid pathways. However, they are actually polymers with fundamental subunits of $\alpha/\beta$ tubulin that gather along their long axis into protofilaments. Generally, twelve to fourteen protofilaments enfold to form a helical cylinder known as MT that serves as a track for intracellular transport carried out by kinesin and dynein family \cite{alberts2008molecular}. Owing to the helical nature, experimentally it has been proved that MTs display wavy trajectories due to the internal organization of protofilaments \cite{gosselin2016complex,howard2001mechanics}. Furthermore, these MTs are surrounded by motor proteins, mitochondria, ribosomes and many other organelles suspended in the cell's cytoplasm that provides a three dimensional (3D) environment to MT \cite{cai2001microtubule,luby1986probing}. Considering polymer like structure of pathways, recently few theoretical models have been proposed to study the steady-state properties of a one dimensional (1D) flexible track immersed in 3D pool of particles \cite{fernandes2019driven, fernandes2017gene, verma2019stochastic}. The key motivation was to analyse the crucial role of ribosomes in the formation of proteins where strands of mRNA are flexible polymers as suggested by experiments. Moreover, \emph{in vitro} and \emph{in vivo} experiments revealed that translation process is greatly influenced by interplay between local concentration of ribosomes, conformation of mRNA and recycling \cite{bakshi2012superresolution}. In this direction, TASEP was coupled to 3D surroundings and the characteristics of the system under the assumption that source and sink in the local neighborhood of entry and exit sites do not show alterations with time were thoroughly analysed. The interaction between transportation of particles and lattice conformation lead to the emergence of new physics in terms of extended shock phase.

The underlying modules that participate in the above explained biological processes are motors and highways. Hence, the variants of TASEP can be studied to reveal interesting physical outcomes of non-equilibrium transportation processes.

Stimulated by such biological processes and features we propose a framework to provide natural means to consider a two particle system moving on a flexible lattice in opposite direction immersed in a 3D pool of infinite particles. The paper is mainly divided into two parts where we first investigate the role of coupling bidirectional TASEP model with 3D environment of diffusing particles. Further we attempt to explore how the particle density interacts with the lattice conformation and regulates the entry rate of particles fluctuating the end-to-end distance of lattice. To explore the overall dynamics of proposed model we compute stationary density profiles, phase diagrams and density dependent distance between two ends of the lattice. We expect that the proposed study might provide an insight to analyze the collective dynamics of out-of-equilibrium transport systems in physical and biological world.

\section{Model}\label{sec2}
Motivated by intracellular transport of two different motor proteins moving in opposite directions namely kinesin and dynein along a polymer like microtubule (MT), we study their bidirectional motion on a flexible lattice with finite number of sites $N$. These motors are denoted by `$+$' and `$-$' in the framework of TASEP where a site can either be empty or occupied by exactly one particle. Moreover, these MTs are structural molecular highways present in the cell's cytoplasm where motors are randomly dispersed. Therefore, a lattice is assumed to be immersed in a three dimensional (3D) pool of infinite diffusing particles. The lattice attains its flexibility when the persistence length $l_p$ is less than $N$  with end-to-end distance $R=\sqrt{2l_pN}$ \cite{fernandes2019driven}. The `$+$' particles hop from left to right, whereas `$-$' particles move in reverse direction on the lattice with unit rate in the bulk when the next neighboring site is empty. For the theoretical study we have assumed the particle length to be equal $\ell=1$. When a `$+$' particle encounters `$-$' particle on the next site or vice versa they swap their position with rate $q$. Both the particles leave the lattice from their respective ends with rate $\beta$. Throughout the paper wherever we use the notation `$\pm$' in the subscript, it means `$+$' and `$-$' particles respectively. The positive and negative particles are injected into the lattice from the two extreme ends $i=1$ and $N$ respectively whenever the target site is empty. The rate at which they enter $\bar{\alpha}_{\pm}$ depends on the local concentration of particles in a sphere of radius $a$ with volume $V_a$ assumed in the neighborhood of two ends, is given by \cite{fernandes2019driven}
\begin{equation}\label{eqn0}
\bar{\alpha}_{\pm}=\dfrac{\alpha_{0}}{V_a}\int\limits_{0}^{2\pi}\int\limits_{0}^{\pi}\int\limits_{0}^{a}c_{\pm}(r,\theta,\phi)r^2\sin{\theta}dr d\theta d\phi.
\end{equation}
where $c_{\pm}(r,\theta,\phi)$ represents the concentration of identical particles present at position $(r,\theta,\phi)$ in a spherical coordinate system and $\alpha_{0}$ is the reaction rate constant. Fig.\ref{fig1} is the schematic illustration of various processes and corresponding rates in the system.

\subsection{Theoretical Description}\label{sec2a}
Biologically motivated, in this section we provide a mathematical support to all the processes seized in the proposed model.
The occupancy of site $i$ is denoted by $\tau^{(i)}_{+}$ and $\tau^{(i)}_{-}$ which takes two values 1 (0) when the site is occupied (empty) with `$+$' and `$-$' particles respectively.
The time evolution equations that govern the dynamics of oppositely moving particles in the bulk ($i=2,3,\hdots,N-1$) are,
\begin{eqnarray}\nonumber
\dfrac{d\langle\tau^{(i)}_+\rangle}{dt}&=&\langle\tau^{(i-1)}_+(1-\tau^{(i)}_+-(1-q)\tau^{(i)}_-)\rangle-\\&&\label{eqn1}\langle\tau^{(i)}_+(1-\tau^{(i+1)}_+-(1-q)\tau^{(i+1)}_-)\rangle,\\\nonumber
\dfrac{d\langle\tau^{(i)}_-\rangle}{dt}&=&\langle\tau^{(i+1)}_+(1-\tau^{(i)}_--(1-q)\tau^{(i)}_+)\rangle-\\\nonumber&&\langle\tau^(i)_-(1-\tau^{(i-1)}_--(1-q)\tau^{(i-1)}_+)\rangle,
\end{eqnarray}
where $\langle\hdots\rangle$ denotes the statistical average.
Similarly, the equations at boundaries $i=1$ and $N$ are expressed as,
\begin{eqnarray}\nonumber
\dfrac{d\langle\tau^{(1)}_+\rangle}{dt}&=&\bar{\alpha}_+\langle(1-\tau^{(1)}_+-\tau^{(1)}_-)\rangle-\\\nonumber&&\langle\tau^{(1)}_+(1-\tau^{(2)}_+-(1-q)\tau^{(2)}_-)\rangle,\\\nonumber
\dfrac{d\langle\tau^{(1)}_-\rangle}{dt}&=&\langle\tau^{(2)}_-(1-\tau^{(1)}_+-(1-q)\tau^{(1)}_-)\rangle-\beta\langle\tau^{(1)}_-\rangle,\hspace{4em}\\\label{eqn2}
\dfrac{d\langle\tau^{(N)}_+\rangle}{dt}&=&\langle\tau^{(N-1)}_+(1-\tau^{(N)}_+-(1-q)\tau^{(N)}_-)\rangle-\beta\langle\tau^{(N)}_+\rangle,\\\nonumber
\dfrac{d\langle\tau^{(N)}_-\rangle}{dt}&=&\bar{\alpha}_-\langle(1-\tau^{(N)}_+-\tau^{(N)}_-)\rangle-\\\nonumber&&\langle\tau^{(N)}_-(1-\tau^{(N-1)}_--(1-q)\tau^{(N-1)}_+)\rangle.
\end{eqnarray}
For $q=1$, equations in the bulk Eq.\ref{eqn1} are decoupled and the system of two distinct particles interact only at the boundaries that we will consider in our present study \cite{evans1995spontaneous,evans1995asymmetric}. In order to evaluate effective entry rate given in Eq.\ref{eqn0}, local concentration $c_{\pm}(r,\theta,\phi)$ is computed by solving a diffusion equation having sink $Q_{\pm}^{out}$ and source $Q_{\pm}^{in}$ terms as follows \cite{fernandes2019driven},
\begin{eqnarray}\label{eqn3}
D\nabla^2c_{\pm}(r,\theta,\phi)=Q_{\pm}^{in}(r,\theta,\phi)-Q_{\pm}^{out}(r,\theta,\phi).
\end{eqnarray}
where $D$ is the diffusion constant of particles in the reservoir. The sink acts as a part of reservoir from which lattice consumes the particles and source provides particles back into the environment. Since, the two distinct particles are moving in opposite directions on the lattice, the sink at $i=1~(N)$ for positive (negative) particles acts as a source for negative (positive) particles. The reaction volumes of source and sink are treated to be equal $Q_{\pm}^{in}=J_{\pm}/V_a$ and $Q_{\pm}^{out}=-J_{\pm}/V_a$ respectively \cite{fernandes2019driven}. This associates the diffusion equation to the lattice current $J_{\pm}$ (induced due to `$+$' and `$-$' particles respectively) which couples the system to TASEP model.
In analogous to calculate the potential in a system with two spherical distribution of charges in electrostatics \cite{eisler1969introduction}, and adoption of similar approach gives the concentration $c_{\pm}(\textbf{x})$ at a distance \textbf{x} from the center of each sphere \cite{fernandes2019driven},
\begin{eqnarray}\label{eqn4}
c_{\pm}(\textbf{x})=
\begin{cases}
\frac{J_{\pm}}{4\pi D|\textbf{x}|} &\quad \text{outside}~ Q^{in}_{\pm},\\
-\frac{J_{\pm}}{4\pi D|\textbf{x}|} &\quad \text{outside}~ Q^{out}_{\pm},\\
\frac{J_{\pm}}{8\pi Da^3}(3a^2-x^2) &\quad \text{inside}~ Q^{in}_{\pm}\\
\frac{J_{\pm}}{8\pi Da^3}(x^2-3a^2) &\quad \text{inside}~ Q^{out}_{\pm}.
\end{cases}
\end{eqnarray}
Using Eq.\ref{eqn4} in Eq.\ref{eqn0} and following ref.\cite{fernandes2019driven}, we obtain the effective entry rates as $\bar{\alpha}_{\pm}=\alpha_{\infty}+J_{\pm}\Gamma$, where $\alpha_{\infty}=\alpha_0c_{\infty}$ ($c_{\infty}$ is the density considered far away from the lattice) that denotes the entry rate of particles considered in standard bidirectional TASEP model with rigid lattice. The parameter $\Gamma$ signifies the recycling strength of particles around the entry site and mainly depends on $R$. Now, when $l_p>\ell$, the conformation of lattice is independent of density of particles and $R$ remains invariant. On the contrary, if $l_p<\ell$ the presence particles affect the lattice conformation that leads to fluctuation in $R$. As a consequence, for the former case recycling strength is constant, whereas, for the latter case it is density dependent \cite{fernandes2019driven}. It is to be noted that all the lengths such as $\ell$, $l_p$, $R$ are in the unit of lattice size. In the proposed model we will explore the interplay between bidirectional transport and lattice conformation under both the cases separately in the upcoming sections.
 \begin{table*}
\begin{center}
\resizebox{\textwidth}{!}{
\begin{tabular}{|c|c|c|c|c|}
  \hline
  \textbf{Phase} & \textbf{Effective entry rates} & \textbf{Bulk Density} & \textbf{Current} ($J_{\pm}$) & \textbf{Phase Region} \\\hline
 LD & $\bar{\alpha}^{LD}_{\pm}=\dfrac{\Gamma-1\sqrt{(1-\Gamma)^2+4\alpha_{\infty}\Gamma}}{2\Gamma}$ & $\bar{\alpha}^{LD}_{\pm}$ & $\bar{\alpha}^{LD}_{\pm}(1-\bar{\alpha}^{LD}_{\pm})$ & $\bar{\alpha}^{LD}_{\pm}<\beta$, $\bar{\alpha}^{LD}_{\pm}<1/2$ \\\hline
  HD & $\bar{\alpha}^{HD}_{\pm}=\alpha_{\infty}+\beta(1-\beta)\Gamma$ & $1-\beta$ & $\beta(1-\beta)$ & $\bar{\alpha}^{HD}_{\pm}>\beta,~\beta<1/2$  \\\hline
  MC & $\bar{\alpha}^{MC}_{\pm}=\alpha_{\infty}+\frac{\Gamma}{4}$ & 1/2 & 1/4 & $\bar{\alpha}^{MC}_{\pm}>1/2,~\beta>1/2$\\
  \hline
\end{tabular}
}
\end{center}
\caption{\label{table}Summary of TASEP results with flexible lattice in form of effective entry rates, bulk density, current and phase boundaries for various phases~\cite{fernandes2019driven} with unit hopping rate in the bulk. Here $\bar{\alpha}_{\pm}$ and $\beta$ denote the entry and removal rate of particles, respectively.}
\end{table*}
\subsection{Mean-field Theory}
In this section we approximate the temporal equations using mean-field theory \cite{derrida1992exact}
that ignores the spatial correlations and factorizes the correlation function into product of their averages,
$\langle\tau^{(i)}_{\pm}\tau^{(i+1)}_{\pm}\rangle=\langle\tau^{(i)}_{\pm}\rangle\langle\tau^{(i+1)}_{\pm}\rangle.$ We denote the average densities of positive and negative particles as,
\begin{eqnarray}\label{eqn5}
\rho_{i}=\langle\tau^{(i)}_+\rangle,~\text{and}~\sigma_{i}=\langle\tau^{(i)}_-\rangle
\end{eqnarray}
respectively. As a result, the equations that determine the particle density for $i=1$ to $N$ reduces to the form,
\begin{eqnarray}\label{eqn6}
\dfrac{d\rho_i}{d t}&=&J_+^{(i-1)}-J_+^{(i)},\\\nonumber
\dfrac{d\sigma_i}{d t}&=&J_-^{(i-1)}-J_-^{(i)},
\end{eqnarray}
where, the current in the bulk $i=2,\hdots,N-1$ is given by,
\begin{eqnarray}\label{eqn7}
J_+^{(i)}&=&\rho_{i}(1-\rho_{i+1}),\\\nonumber
J_-^{(i)}&=&\sigma_{i+1}(1-\sigma_{i}),
\end{eqnarray}
and at the boundaries $i=1$ and $N$ is,
\begin{eqnarray}\nonumber
J_+^{(1)}&=&\bar{\alpha}_+(1-\rho_1-\sigma_1),\\\nonumber
J_-^{(1)}&=&\beta\sigma_1,\\\label{eqn8}
J_+^{(N)}&=&\beta\rho_N,\\\nonumber
J_-^{(N)}&=&\bar{\alpha}_-(1-\rho_N-\sigma_N).
\end{eqnarray}
Clearly, Eq.\ref{eqn6} shows that in steady state the current is constant throughout the lattice and hence superscript $i$ can be dropped in the above expressions. Further, utilizing these equations we study the dynamics of the proposed system under the influence of important controlling parameters at the boundaries i.e. entry and removal rates.

\section{Results and Discussions}
To explore the two different scenarios mentioned in section \ref{sec2a}, we study the emerging dynamics in $(\alpha_{\infty},\beta)$ space. It is anticipated that when $R$ is invariant, the topology of phase diagram remains conserved and is altered quantitatively. Whereas, for the case when particle density interacts with the lattice conformation and $R$ varies, we observe the non-trivial effect on the topology of phase diagram. In this section we investigate the steady-state properties of the system in detail under the influence of boundary controlling parameters for $N=25$ and when $R$ is density dependent the other required parameters are $\ell=1$, $l_p=0.1$, $\frac{\alpha_0}{4\pi Da}=2.25$ and $a=1.5$ \cite{fernandes2019driven}.

\subsection{Flexible lattice with invariant end-to-end distance $(l_p>l)$}
In this section, we investigate the case when the transport of distinct particles on a flexible lattice do not alter the end-to-end distance $R$ and $\Gamma$ turns out to be a negative constant. For more details to know how recycling strength $\Gamma$ is calculated we refer the reader to ref.\cite{fernandes2019driven}. Now, to analytically solve the governing mean-field equations interacting only at the boundaries, we define modified entry rate for `$+$' and `$-$' particles as $\alpha_+^E$ and $\alpha_-^E$ respectively \cite{evans1995spontaneous,evans1995asymmetric},
\begin{eqnarray}\label{eqn9}
\alpha_+^E(1-\rho_1)&=&\bar{\alpha}_+(1-\rho_1-\sigma_1),\\\label{eqn10}
\alpha_-^E(1-\sigma_N)&=&\bar{\alpha}_-(1-\rho_N-\sigma_N),
\end{eqnarray}
which easily reduces to
\begin{eqnarray}\label{eqn11}
\alpha_+^E=\dfrac{J_+}{J_+/\bar{\alpha}_++J_-/\beta},\\\label{eqn12}
\alpha_-^E=\dfrac{J_-}{J_-/\bar{\alpha}_-+J_+/\beta}.
\end{eqnarray}
Note that in contrast to bidirectional transport on rigid lattice, here $\bar{\alpha}_+$ and $\bar{\alpha}_-$ will vary according to phases as summarized in table \ref{table} \cite{fernandes2019driven}.
By the current constancy condition current in the lattice reduces to,
\begin{eqnarray}\label{eqn13}
J_+=\alpha_E^+(1-\rho_1)=\rho_1(1-\rho_2)=\hdots=\beta\rho_N,\\\label{eqn14}
J_-=\beta\sigma_1=(1-\rho_1)\rho_2=\hdots=\alpha_-^E(1-\sigma_1).
\end{eqnarray}
The input and output of particles from the boundaries regulate the state of transport on the flexible lattice, so it becomes significant to analyse the stationary phase diagram in  $(\alpha_{\infty},\beta)$ plane. For $\Gamma\rightarrow 0$, the system reveals the phenomenon of SSB that exhibits unequal current induced due to distinct particles and results in two asymmetric phases; high density-low density (HD-LD) and asymmetric low density (L-L) phase. In addition there are two symmetric phases where current is equal in the lattice for both the particles; low density-low density (LD-LD) and maximal current-maximal current (MC-MC) phase \cite{evans1995spontaneous,evans1995asymmetric}. Now, we investigate the existence of possible phases when $\Gamma$ is a non-zero quantity.\\
\begin{figure*}[ht!]
            \centering
\subfigure[\label{fig2a}]{\includegraphics[width=.35\textwidth,height=6cm]{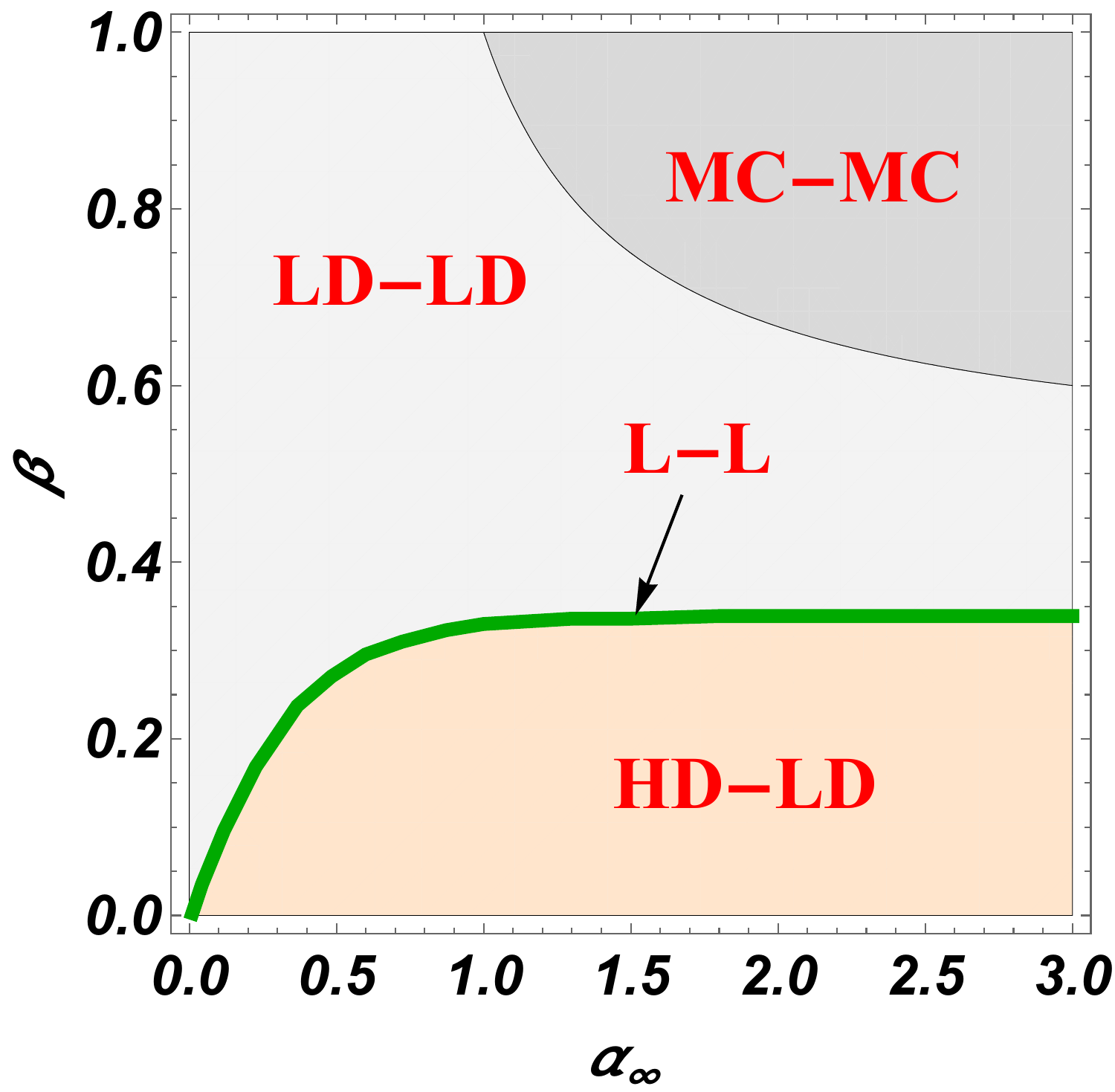}}\hspace{7mm}
\subfigure[\label{fig2b}]{\includegraphics[width=.35\textwidth,height=6cm]{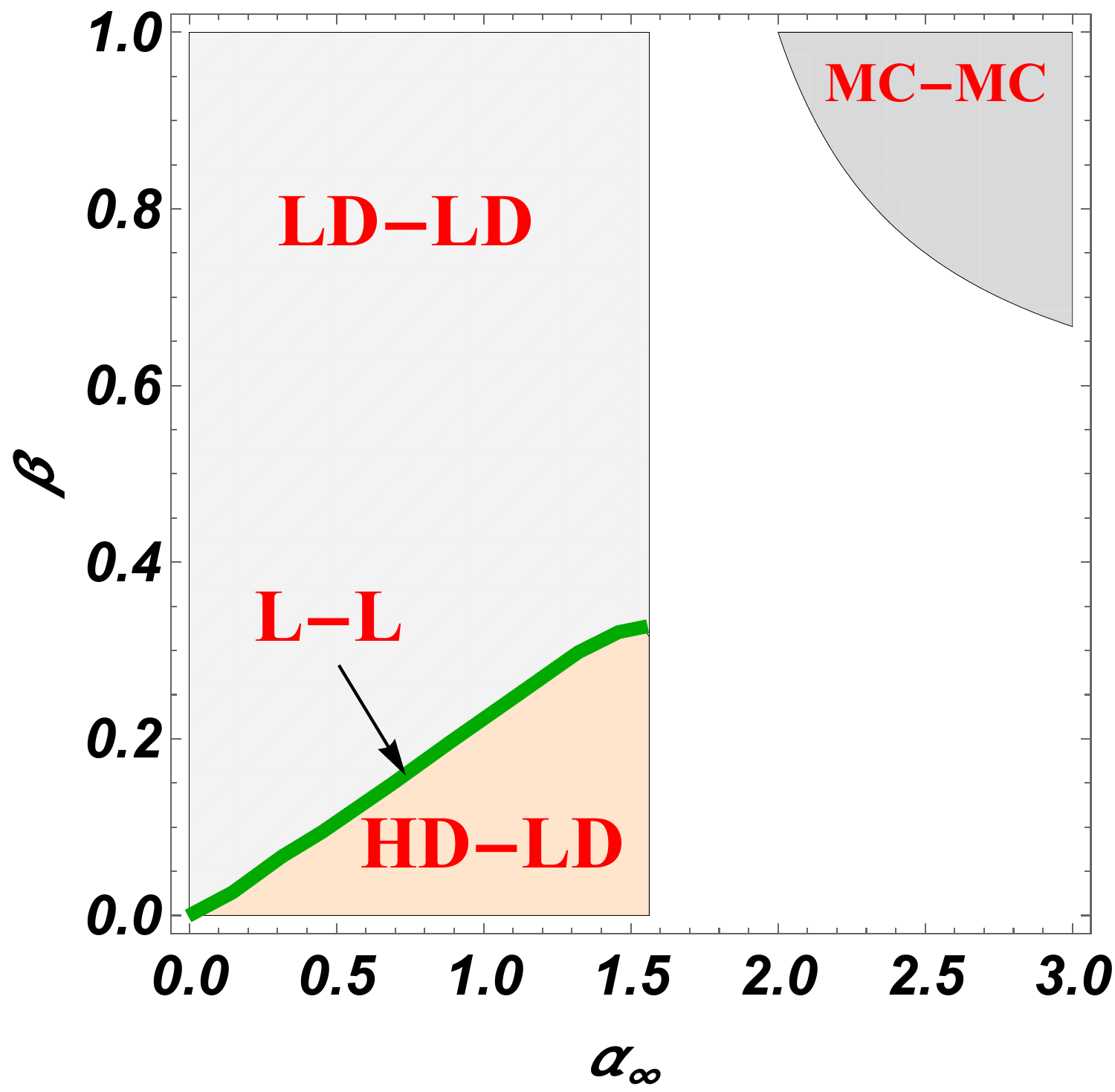}}
\caption{\label{fig2} Phase diagrams for (a) $\Gamma=0$ (b) $\Gamma=-4$ with $N = 25$. Green line represents a narrow region of asymmetric low density phase ($L-L$). In (b) the unshaded portion represents the unknown phase for $\alpha_{\infty}<\frac{2-\Gamma}{4}$.}
\end{figure*}

\subsubsection{Symmetric Phases}
Here, we discuss the properties of system when the density and current induced is equal for both type of particles i.e. $J_+=J_-$. This implies that entry rate of both the particles is equal $\bar{\alpha}_+=\bar{\alpha}_-$ for a fixed $\Gamma$ and $\alpha_{\infty}$. Further solving Eqs.\ref{eqn11} and \ref{eqn12} yields,
\begin{equation}\label{eqn15}
\alpha^E_+=\alpha^E_-=\dfrac{\bar{\alpha}_+\beta}{\bar{\alpha}_++\beta}=\alpha^E~(\text{say}).
\end{equation}
Utilizing Eq.\ref{eqn15} we analyse the various observed symmetric phases in the proposed system.
\begin{enumerate}
  \item \emph{LD-LD} Phase:\\ \\
 Low density symmetric phase exists when the density of both the particles is entry dominated and bulk density is given by $\alpha^E$. The entry rate that governs the dynamics is $\bar{\alpha}_+=\bar{\alpha}^{LD}_+$ given in table \ref{table}.
  The conditions for existence of this phase are,
\begin{equation}\label{eqn16}
  \alpha^E<\min\{\beta,1/2\}.
  \end{equation}
  where $\alpha^E$ is calculated by utilizing Eq.\ref{eqn15}. This further reads out that this phase is possible when
  \begin{equation}\label{eqn17}
  \alpha_{\infty}<\beta\Big(\dfrac{2\Gamma(\beta-1)+1}{(2\beta-1)^2}\Big),\quad \beta>0
   \end{equation}
   provided $\alpha_{\infty}<\dfrac{2-\Gamma}{4}$. We plot the same for $\Gamma=-4$ in Fig.\ref{fig2b}. In the limit $\Gamma\rightarrow 0$ the expressions reduce to the case of bidirectional transport on rigid lattice \cite{evans1995asymmetric} (see Fig.\ref{fig2a}).
  \item \emph{MC-MC} Phase:\\ \\
  In this phase the uniform bulk density is $1/2$ for both the species and $\bar{\alpha}_+=\bar{\alpha}_+^{MC}$ directs the entry rate of particles as given in table \ref{table}. The conditions that determine this phase are,
  \begin{equation}\label{eqn18}
  \alpha^E>1/2,\quad \beta>1/2,
  \end{equation}
  which on using Eq.\ref{eqn15} gives
  \begin{equation}\label{eqn19}
  \alpha_{\infty}>\dfrac{\beta}{2\beta-1}-\dfrac{\Gamma}{4},\quad\beta>1/2.
  \end{equation}
  We observed that as $\Gamma$ decreases MC-MC phase shrinks in comparison to that obtained for $\Gamma=0$ as delineated in Fig.\ref{fig2}.
  \item \emph{HD-HD} Phase:\\ \\
  The properties of this phase will be exit dominated that satisfies the condition,
  \begin{equation}\label{eqn20}
  \alpha^E>\beta,\quad \beta<1/2.
  \end{equation}
  \begin{figure}[ht!]
\centering
 \hspace{0cm} \includegraphics[trim = 180 250 190 260,width=.27\textwidth,height=5cm]{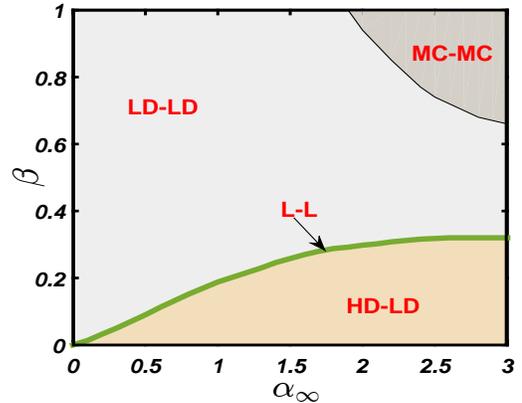}
            \caption{\label{fig3} Phase diagram for $\Gamma=-4$ with $N=25$ when $R$ is independent of density of particles. This diagram fills the whole $(\alpha_{\infty},\beta)$ plane displaying the phase regime where analytic approach breakdown in Fig.\ref{fig2b}.}
\end{figure}
             \begin{figure*}[ht!]
             \centering
\hspace*{1cm}\subfigure[]{\includegraphics[trim = 180 230 100 240,width=.4\textwidth,height=6cm]{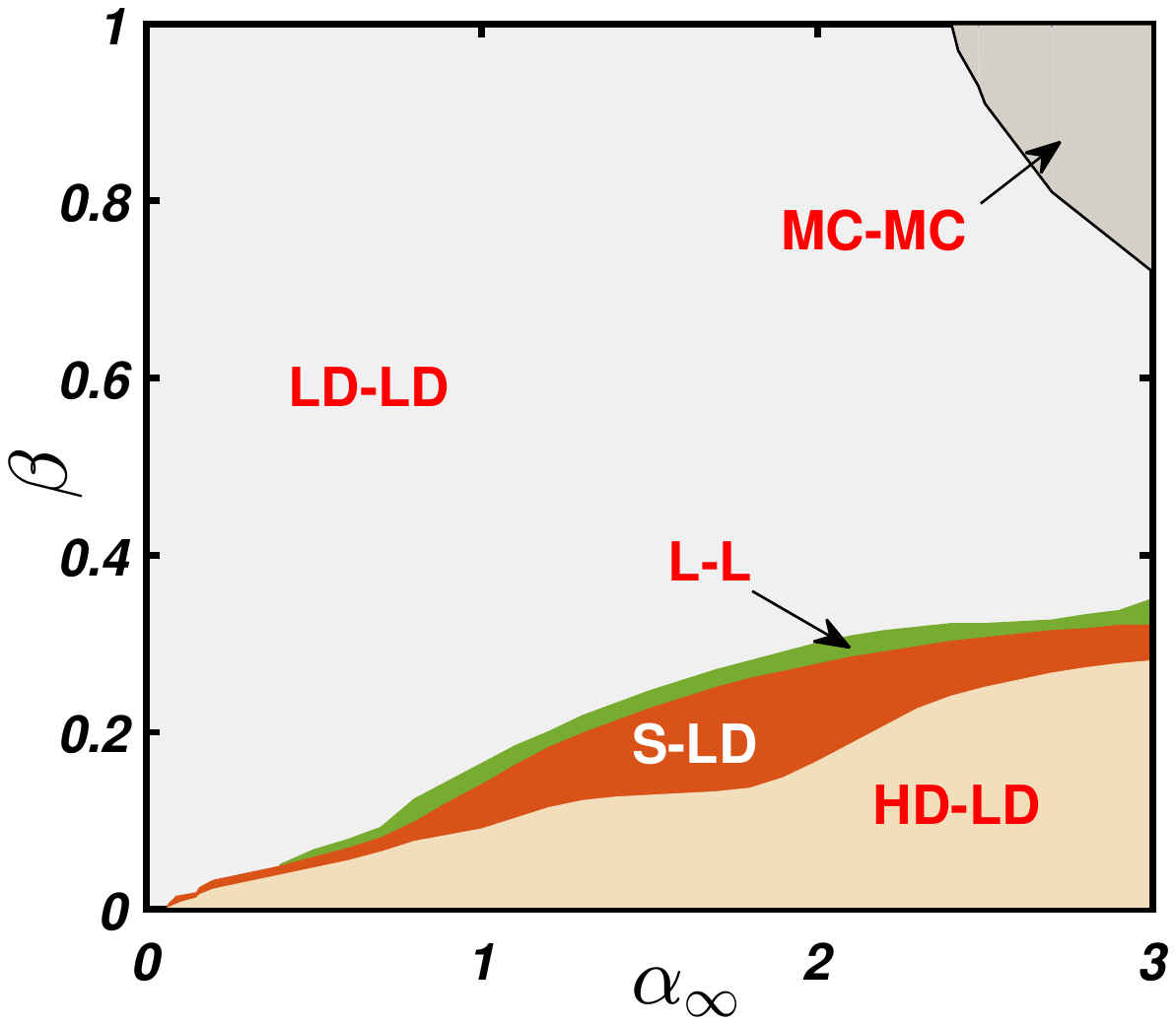}}
\hspace*{1cm}\subfigure[]{\includegraphics[trim = 180 230 100 240,width=.4\textwidth,height=6cm]{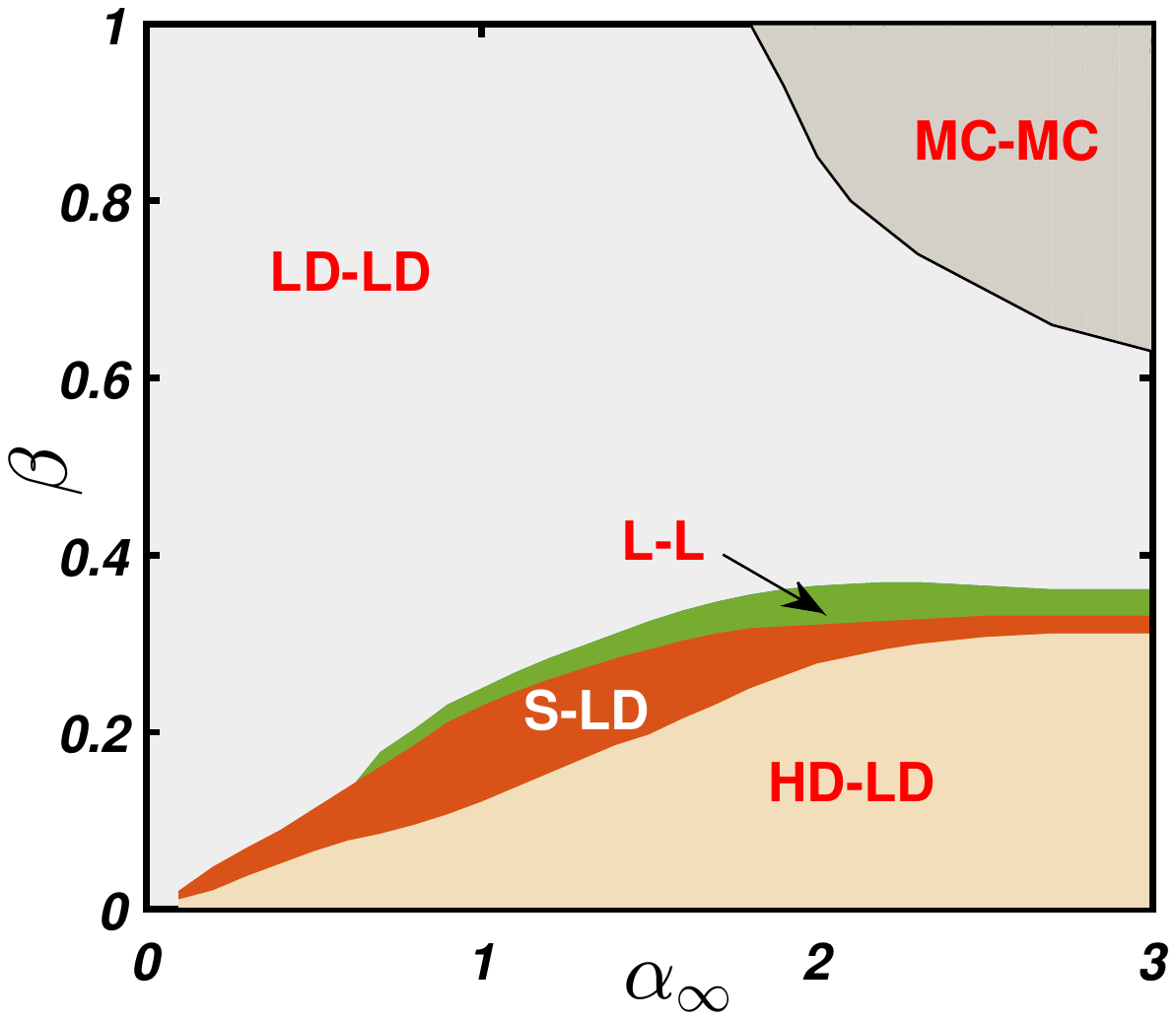}}
\caption{\label{fig4} Stationary phase diagrams for (a) $\Gamma=-2.5$ (b) $\Gamma=-5$ with $N=25$, $\ell=1$, $l_p=0.1$, and $a=1.5$ when $R$ varies with density of particles on the lattice.}
\end{figure*}
The average density will be $1-\beta$ for both the particles and hence the total density will exceed 1 which is impossible. This discards the possibility of existence of HD-HD phase for the proposed system.
  \item \emph{S-S} (Shock Symmetric Phase):\\ \\
   In this phase both the particles exhibit discontinuous connectivity between LD and HD regions. For both the particles a portion of density will be in HD phase that is impossible as already discussed for HD-HD case. Hence, this phase cease to exist for any value of $(\alpha_{\infty},\beta)$. The reverse possibility of shock from HD to LD has been already discarded \cite{gupta2014asymmetric}.
\end{enumerate}

\subsubsection{Asymmetric Phases}
The interaction of distinct particles at the boundaries affects the symmetry of the system, resulting in the phenomenon of SSB and existence of asymmetric phases. In this phase, two types of particles generally exhibit unequal density and hence unequal current i.e. $J_+\neq J_-$ in the lattice. Here, we investigate the conditions for existence of possible asymmetric phases for the proposed system.
\begin{enumerate}
  \item \emph{L-L} (Asymmetric low density) Phase:\\ \\
  Since, this phase is entry dominated for both the particles, the current in the lattice is given by,
  \begin{equation}\label{eqn21}
  J_+=\alpha_+^E(1-\alpha_+^E),\quad J_-=\alpha_-^E(1-\alpha_-^E)
  \end{equation}
  with bulk densities $\alpha_+^E$ and $\alpha_-^E$.
  Utilizing Eqs.\ref{eqn11} and \ref{eqn12} with $\bar{\alpha}_+=\bar{\alpha}_+^{LD}$ and $\bar{\alpha}_-=\bar{\alpha}_-^{LD}$ leads to,
  \begin{eqnarray}\label{eqn22}
  \alpha_+^E=1+\bar{\alpha}^{LD}_+-\sqrt{(1-(\bar{\alpha}^{LD}_+)^2-\dfrac{4\bar{\alpha}^{LD}_+\alpha^E_-(\alpha^E_--1)}{\beta})},\hspace{2em}\\\label{eqn23}
  \alpha_-^E=1+\bar{\alpha}^{LD}_--\sqrt{(1-(\bar{\alpha}^{LD}_-)^2-\dfrac{4\bar{\alpha}^{LD}_-\alpha^E_+(\alpha^E_+-1)}{\beta})}.\hspace{2em}
  \end{eqnarray}
  The conditions that govern the existence of this phase are,
  \begin{eqnarray}\label{eqn24}
  \alpha_+^E,~\alpha_-^E<\min\{\beta,1/2\}.
   \end{eqnarray}
   \begin{figure*}[ht!]
\centering
\subfigure[]{\includegraphics[trim = 180 240 100 240,width=.31\textwidth,height=4.5cm]{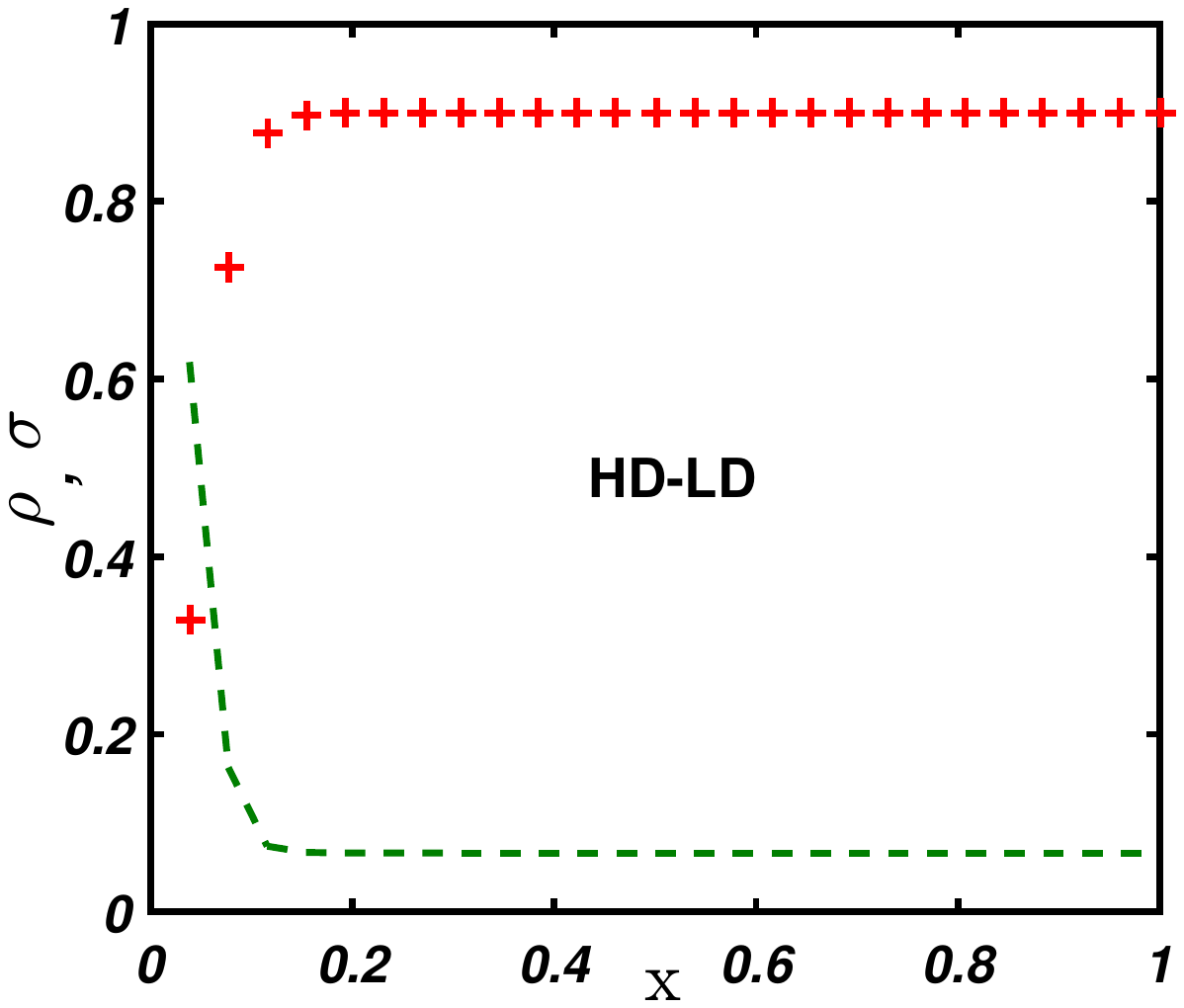}}\hfill
\subfigure[]{\includegraphics[trim = 180 240 100 240,width=.31\textwidth,height=4.5cm]{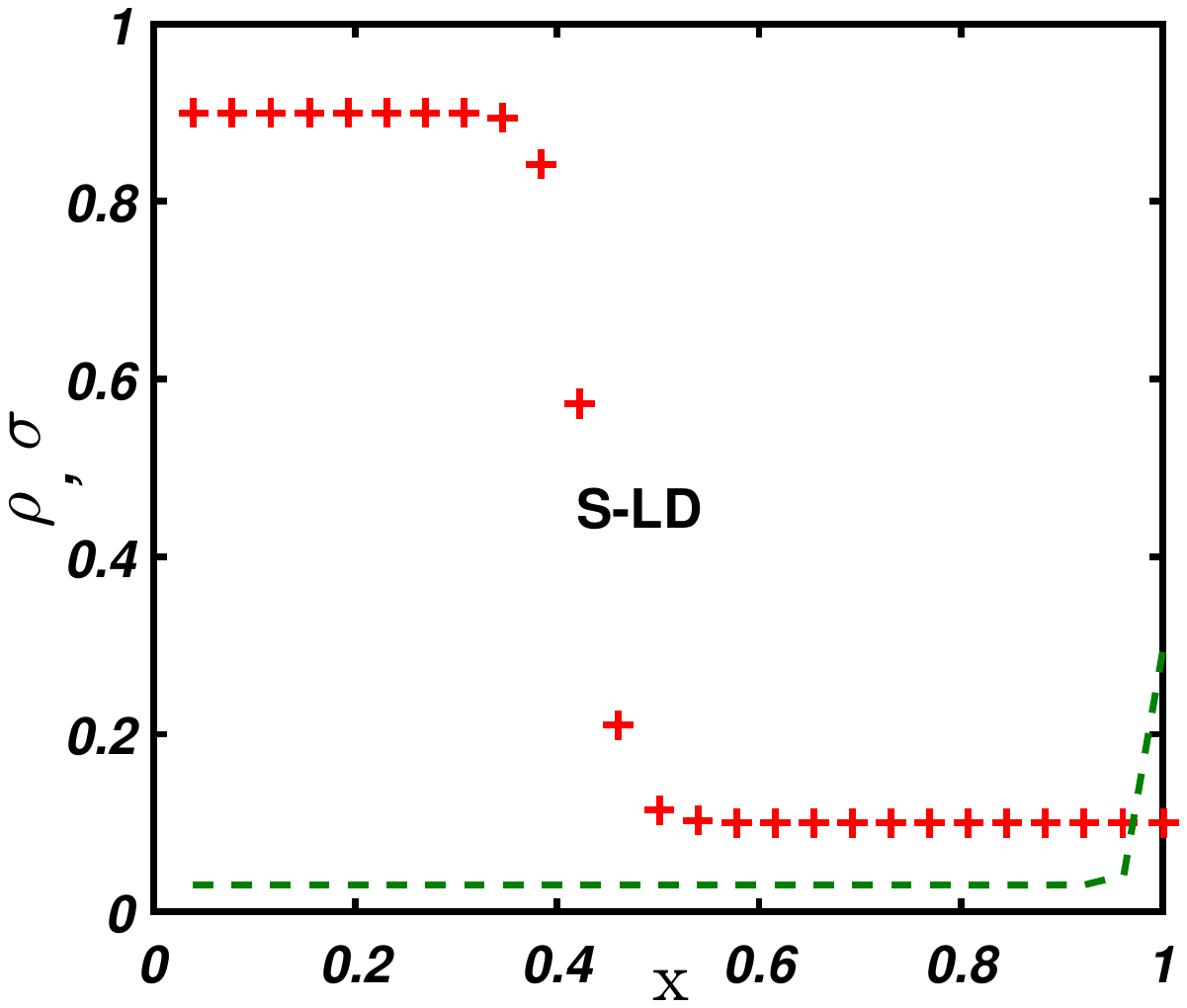}}\hfill
\subfigure[]{\includegraphics[trim = 180 240 100 240,width=.31\textwidth,height=4.5cm]{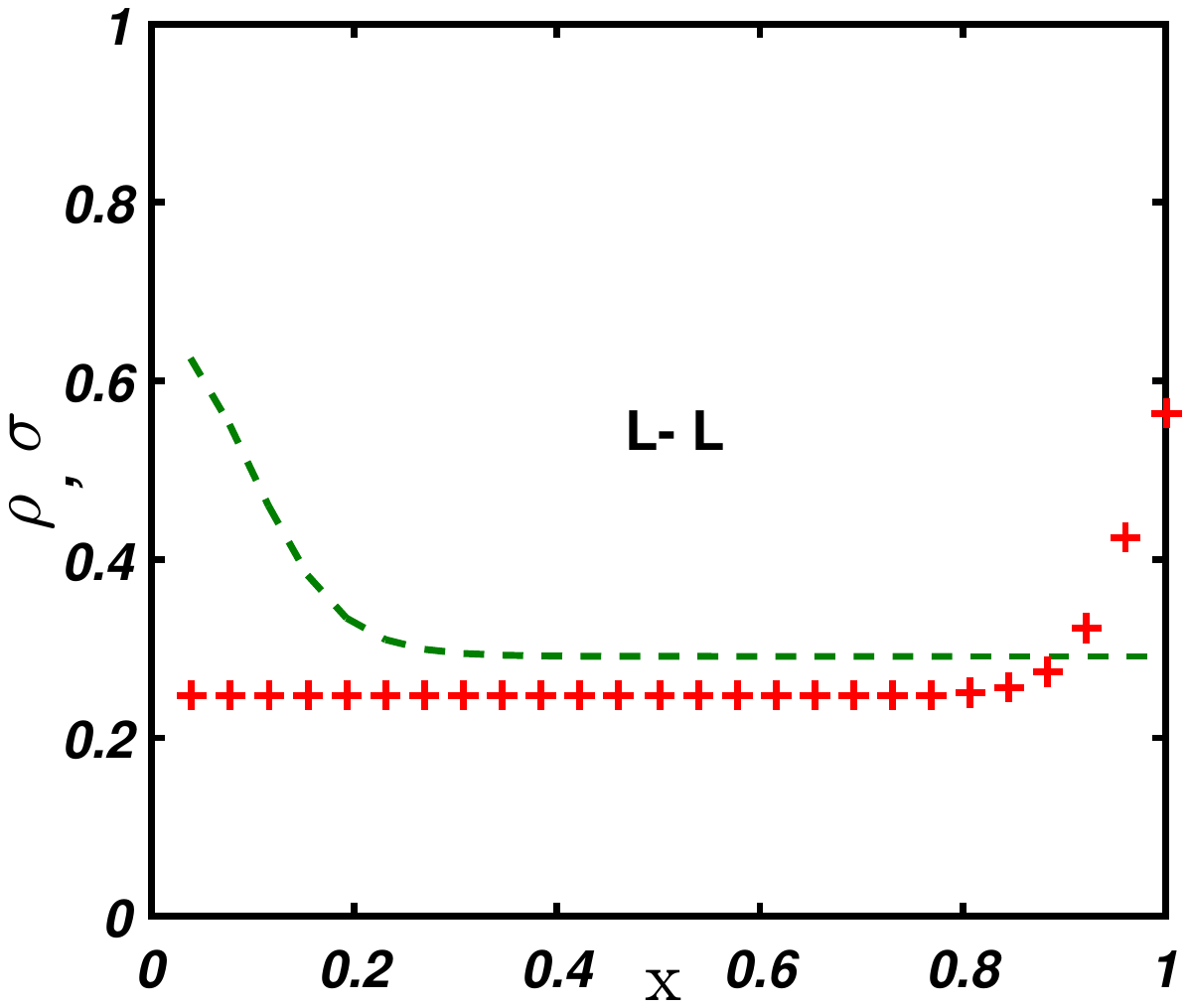}}\\
\subfigure[]{\includegraphics[trim = 180 240 100 240,width=.31\textwidth,height=4.5cm]{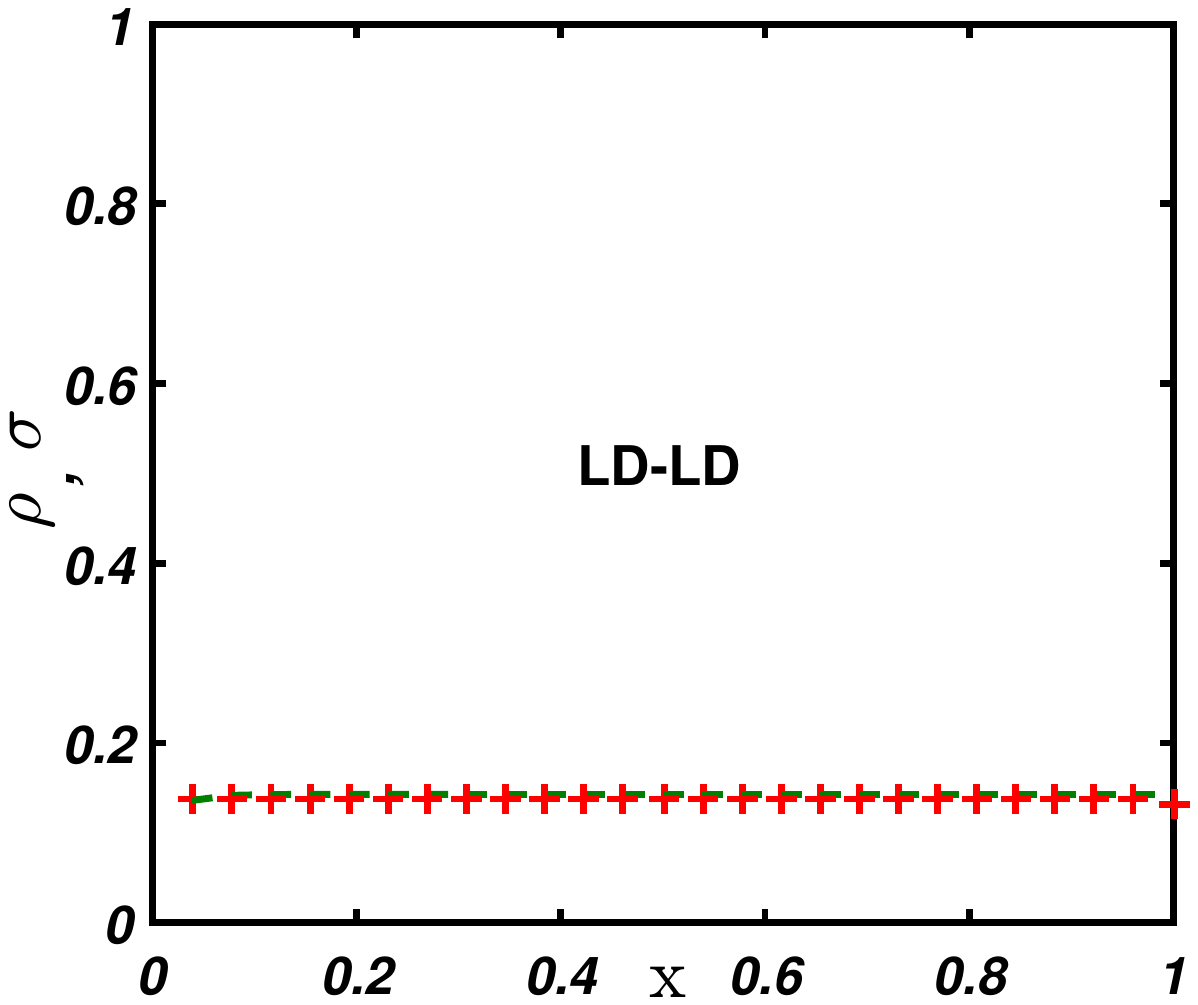}}
\hspace*{0.5cm}\subfigure[]{\includegraphics[trim = 180 240 100 240,width=.31\textwidth,height=4.5cm]{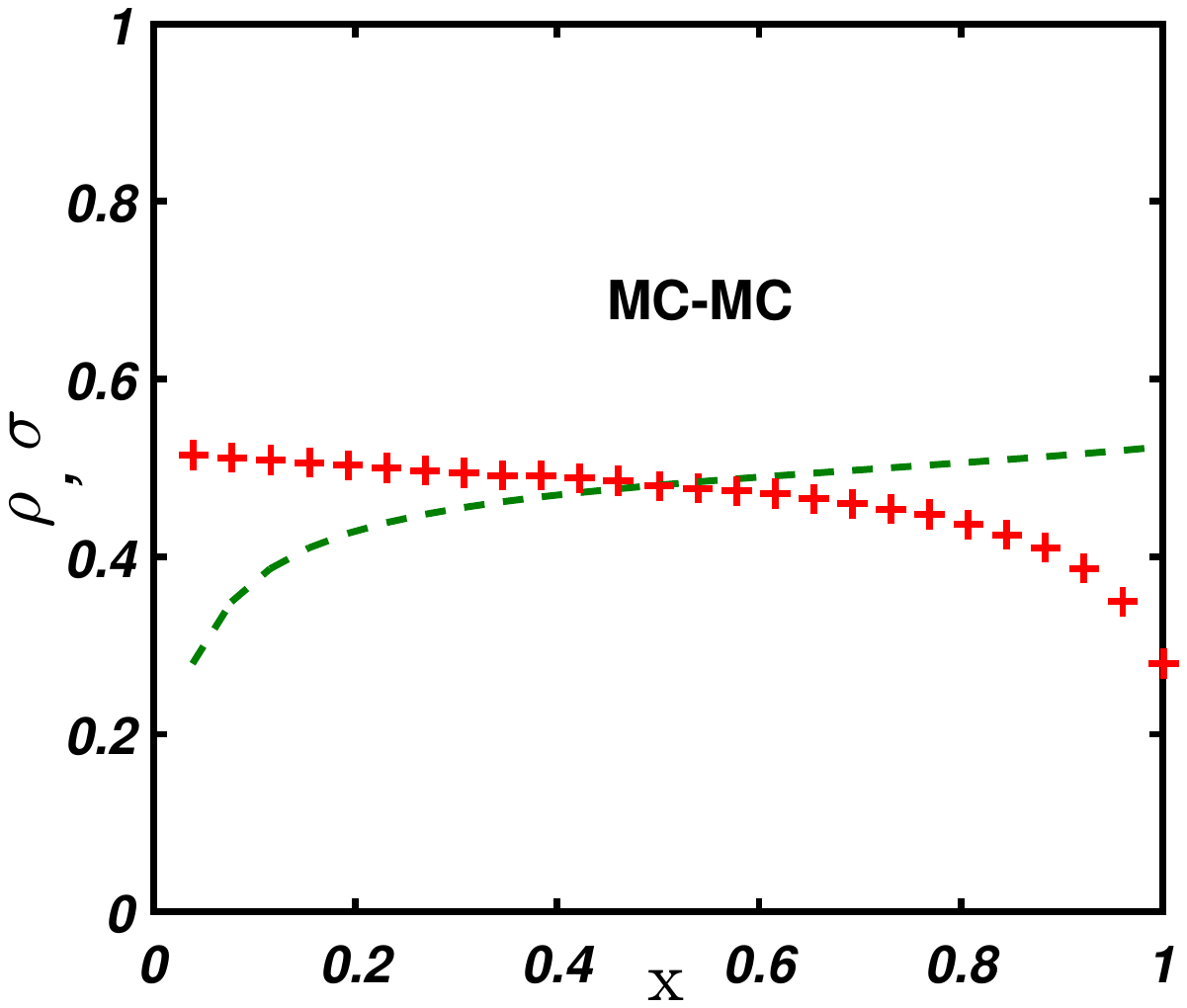}}
\caption{\label{fig5} Typical density profiles of (a) HD-LD (b) S-LD (c) L-L (d) LD-LD (e) MC-MC phases for $\Gamma=-2.5$ and $(\alpha_{\infty}, \beta)$=(2, 0.1), (0.5, 0.1), (2, 0.33), (0.5,0.9), (2,0.9). In all figures other parameter values are $N=25$, $\ell=1$, $l_p=0.1$, and $a=1.5$.}
\end{figure*}
for which using Eqs.\ref{eqn22}-\ref{eqn23} provides this phase regime as shown in Fig.\ref{fig2b}. Clearly, for $\Gamma\rightarrow 0$ the expressions converges to the case when there is no flexibility in the lattice \cite{evans1995asymmetric}.
  \item \emph{HD-LD} Phase:\\ \\
   Without any loss of generality, we assume positive particles to be in HD phase, whereas negative particles in LD phase, thus,
   $$J_+=\beta(1-\beta),\quad J_-=\alpha^E_-(1-\alpha_-^E).$$
   with particle densities $(1-\beta)$ and $\bar{\alpha}_-^{LD}$ respectively. This phase appears when,
   \begin{eqnarray}\label{eqn25}
  \alpha_+^E>\beta,\quad\alpha_-^E<\beta,\quad\beta<1/2.
  \end{eqnarray}
  Using the corresponding effective entry rates of `$+$' and `$-$' particles in Eqs.\ref{eqn11} and \ref{eqn12} yields,
  \begin{eqnarray}\label{eqn26}
  \alpha_+^E=\dfrac{J_+}{J_+/\bar{\alpha}_+^{HD}+J_-/\beta},\\\label{eqn27}
\alpha_-^E=\dfrac{J_-}{J_-/\bar{\alpha}_-^{LD}+J_+/\beta},
\end{eqnarray}
 and results in
 \begin{equation}\label{eqn28}
 \alpha_-^E=\dfrac{1}{2}(1+\bar{\alpha}_-^{LD}-\sqrt{(1+\bar{\alpha}_-^{LD})^2-4\bar{\alpha}_-^{LD}\beta})
  \end{equation}
which is further utilized to compute $\alpha_+^E$. Exploiting the conditions given in Eq.\ref{eqn25} we can derive this phase region as shown in Fig.\ref{fig2b}.
\item \emph{MC-LD} Phase:\\ \\
 Without any loss of generality, assuming positive and negative particles to exhibit MC and LD phase respectively,
the governing criteria for the existence of this phase is,
\begin{eqnarray}\label{con}
\alpha^E_+>1/2,\quad\alpha^E_-<1/2.
\end{eqnarray}
The current in the lattice due to `$+$' and `$-$' particles is given by,
\begin{equation}
J_+=1/4,\quad J_-=\alpha^E_-(1-\alpha^E_-)
\end{equation}
with bulk densities 1/2 and $\alpha^E_-$ respectively. Substituting the corresponding entry rates of `$+$' and `$-$' particles, $\bar{\alpha}_+=\bar{\alpha}_+^{MC}$ and $\bar{\alpha}_-=\bar{\alpha}_+^{LD}$ from table.\ref{table} in Eq.\ref{eqn12} gives,
 \begin{eqnarray}
 \alpha_-^E&=&\dfrac{1}{2}\Big(1+\bar{\alpha}_{LD}-\sqrt{\dfrac{\bar{\alpha}_{LD}}{\beta}+(1-\bar{\alpha}_{LD})^2}\Big)
 \end{eqnarray}
 which is further substituted in Eq.\ref{eqn11} to obtain $\alpha^E_+$.
 Utilizing the conditions given in Eq.\ref{con}, one can see that they are not satisfied for any value of entry and removal rate. Hence, the existence of this phase is discarded.
  \item \emph{S-LD} (Shock-Low density) Phase:\\ \\
  In this phase, we suppose positive particles to display shock phase, while negative particles to exhibit LD phase. Hence, the entry rate for both the particles is governed by $\bar{\alpha}_{\pm}^{LD}$ that results in similar expressions as obtained in Eqs.\ref{eqn22}-\ref{eqn23}. The conditions for the existence of this phase are,
  \begin{eqnarray}\label{eqn29}
  \alpha^E_+<\beta,\quad\alpha^E_+>\beta,\\\nonumber
  \alpha_-^E<\beta,\quad\beta<1/2.
  \end{eqnarray}
  Above inequality cannot be satisfied by any value of ($\alpha_{\infty},\beta$) leading to non-existence of this phase.
  It is worth to mention that asymmetric HD-HD, S-HD, MC-HD, MC-S, S-S phases cease to exist because in these phases the density of particles on the lattice exceeds 1 which is impossible.
\end{enumerate}
\textbf{Breakdown of analytical approach-} One of our main aim is to explore the role of non-zero $\Gamma$ on the topology of phase diagram. In the above discussion, we analytically reproduced the phase schema for $\Gamma\rightarrow 0$ that retrieves bidirectional transport process on a rigid lattice as shown in Fig.\ref{fig2a}. For $\Gamma<0$, we try to analyse the phase diagram utilizing the obtained analytic expressions. Fig.\ref{fig2b} displays the phase diagram for $\Gamma=-4$ from where it is clear that these results do not explore the complete phase region. Theoretical predictions are able to provide the existence of MC-MC, LD-LD, HD-LD and L-L phase regimes. However, the analytic approach is not able to investigate the possible phase beyond $\alpha_{\infty}>\dfrac{2-\Gamma}{4}$ excluding those values of $\alpha_{\infty}$ and $\beta$ for which MC-MC region exists (unshaded portion in Fig.\ref{fig2b}). It is possible that within this range either the existing phases continue to exist or a new dynamics in terms of S-LD and MC-LD phase might emerge. In this direction, to overcome the limitations of analytic approach we move towards utilizing the continuum mean-field approach.\\ \\
\emph{Continuum mean-field approximation-} To scrutinize the stationary phase diagram, we utilize the framework of governing master equations Eq.\ref{eqn6} under continuum mean-field approximation in the thermodynamic limit. For this we consider a lattice constant $\epsilon=1/N$ where $N \rightarrow \infty$ and rescale spatial and temporal variables on the continuum scale as $t\rightarrow t/N$ and $x=i\epsilon$, respectively. Defining $\rho_i\equiv\rho(x,t)$, $\sigma_i\equiv\sigma(x,t)$ and performing Taylor series expansion by retaining terms upto second order, Eqs.\ref{eqn6} reduces to the form,
\begin{eqnarray}\label{eqn30}
\dfrac{\partial}{\partial t}\begin{pmatrix}
\rho\\
\sigma
\end{pmatrix}+\dfrac{\partial }{\partial x}\begin{pmatrix}
-\dfrac{\epsilon}{2}\dfrac{\partial \rho}{\partial x}+\rho(1-\rho)\\
-\dfrac{\epsilon}{2}\dfrac{\partial \sigma}{\partial x}+\sigma(1-\sigma)
\end{pmatrix}=0.
\end{eqnarray}
The suitable boundary conditions are supplied using Eqs.\ref{eqn6} and \ref{eqn8} to solve the above equations numerically utilizing finite-difference scheme as discussed in section \ref{appendix}.
We observed that for a non-zero $\Gamma$, the four phases obtained for $\Gamma\rightarrow 0$ remain intact but the topology of the phase diagram alters quantitatively. With an decrease in $\Gamma$, the effective entry rate of particles reduces that decreases the supply of particles into the lattice. As a consequence, the LD-LD region expands with shrink in HD-LD and MC-MC phase as shown for $\Gamma=-4$ in Fig.(\ref{fig3}). Furthermore, the phase region obtained numerically agrees well to that portion of phase diagram computed analytically for $\alpha_{\infty}<\dfrac{2-\Gamma}{4}$ . This even validates that the numerical scheme described in section \ref{appendix} works well to describe the properties of the system.
\begin{figure}
\includegraphics[trim = 140 240 130 240,width=.4\textwidth,height=7cm]{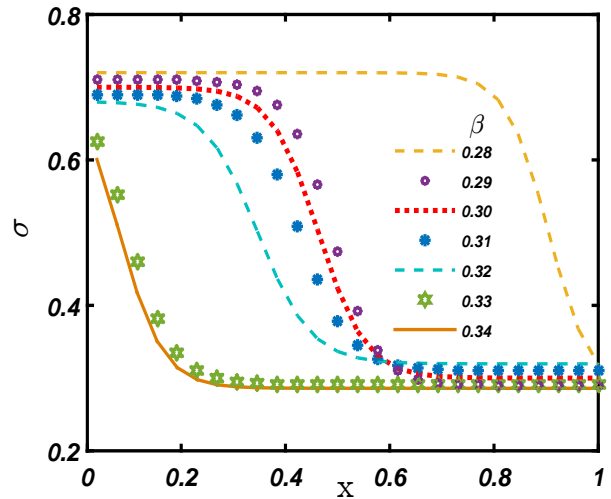}
\caption{\label{fig6} Phase transitions from HD$\rightarrow$S$\rightarrow$LD for particles moving right to left on lattice for $\Gamma=-2.5$ and $N=25$ for $\alpha=2$ with respect to $\beta$. Other parameter values are $\ell=1$, $l_p=0.1$, and $a=1.5$.}
\end{figure}
\subsection{Effect of particle density on end-to-end distance of the lattice $(l_p<\ell)$}
In this section, we investigate the scenario when the transportation of particles govern the global conformation of lattice. This situation persists when $l_p\leq\ell/2$ and the presence of a particle flattens the area of lattice corresponding to it's footprint. As a consequence the density of particles on the lattice affects the end-to-end distance $R$ which further regulates $\Gamma$ that influences the entry rate of particles \cite{fernandes2019driven}.
The density dependent end-to-end distance is given by $R=R_0F$, where $R_0=\sqrt{2l_pN}$ is the distance between two ends when the lattice is empty and
\begin{equation}\label{eqn13}
F=\sqrt{1+\delta\ell\Big(\dfrac{\ell}{2l_p}-1\Big)},
\end{equation}
is the flattening parameter of lattice due to total number of particles $\delta$ (the sum of average densities of positive and negative particles) present on whole lattice.
Since, the effective entry rate of `$+$' and `$-$' particles depend on the recycling of identical particles around their respective entry sites ($i=1$ and $i=N$), the corresponding recycling strength is denoted as $\Gamma_{+}$ and $\Gamma_-$. Depending on the distance between two spheres ($d=R/2a$) in the neighborhood of two extreme ends of the lattice, the strengths are obtained as $\Gamma_{\pm}=\Gamma+\dfrac{\alpha_{0}}{4\pi DR_0}\Big(\dfrac{(1-F_{\pm})}{F_{\pm}}\Big),~~ d \geq 1$ and  $\Gamma_{\pm}=\Gamma+\dfrac{\alpha_{0}d_0^3}{4\pi Da}\Big(\dfrac{3}{2}(F_{\pm}-1)-\dfrac{1}{5}d_0^2({F}_{\pm}^3-1)\Big),~~  d< 1$, where $d_0=R_0/2a$; $F_{+}$ and $F_{-}$ denote the flattening parameter of lattice due to `$+$' and `$-$' particles obtained by replacing $\delta$ with $\rho$ and $\sigma$ respectively in Eq.\ref{eqn13}. Thus, the effective entry rates take the form $\bar{\alpha}_{\pm}=\alpha_{\infty}+J_{\pm}\Gamma_{\pm}$ that are substituted in Eq.\ref{eqn8} to obtain the boundary conditions that are further supplied to Eq.\ref{eqn30} to obtain the steady state properties.\\
\begin{figure*}[ht!]
\centering
\hspace{0cm} \subfigure[\label{fig7b}]{\includegraphics[trim = 180 230 95 240,width=.28\textwidth,height=5cm]{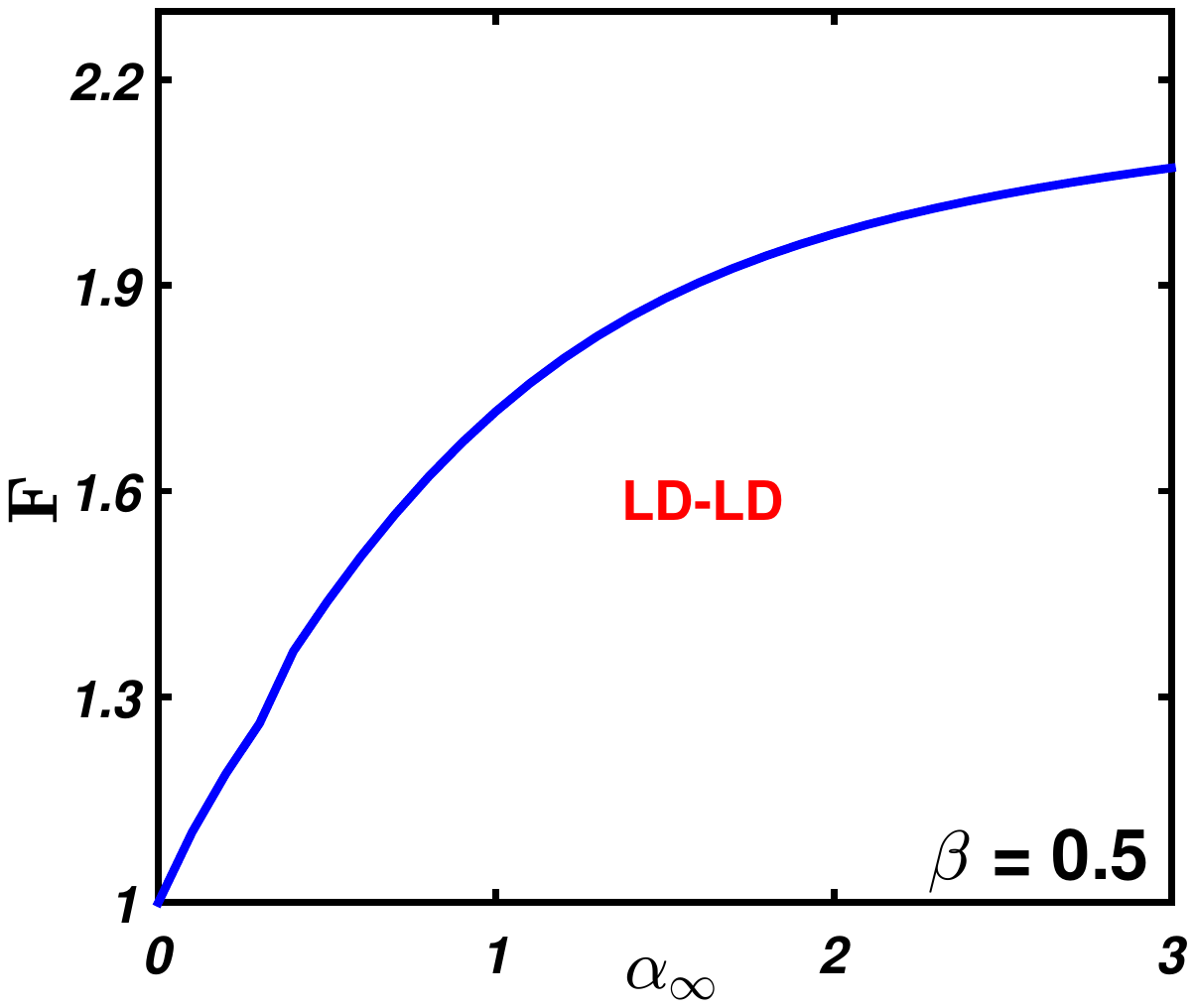}}
\hspace{0cm} \subfigure[\label{fig7c}]{\includegraphics[trim = 180 230 100 240,width=.28\textwidth,height=5cm]{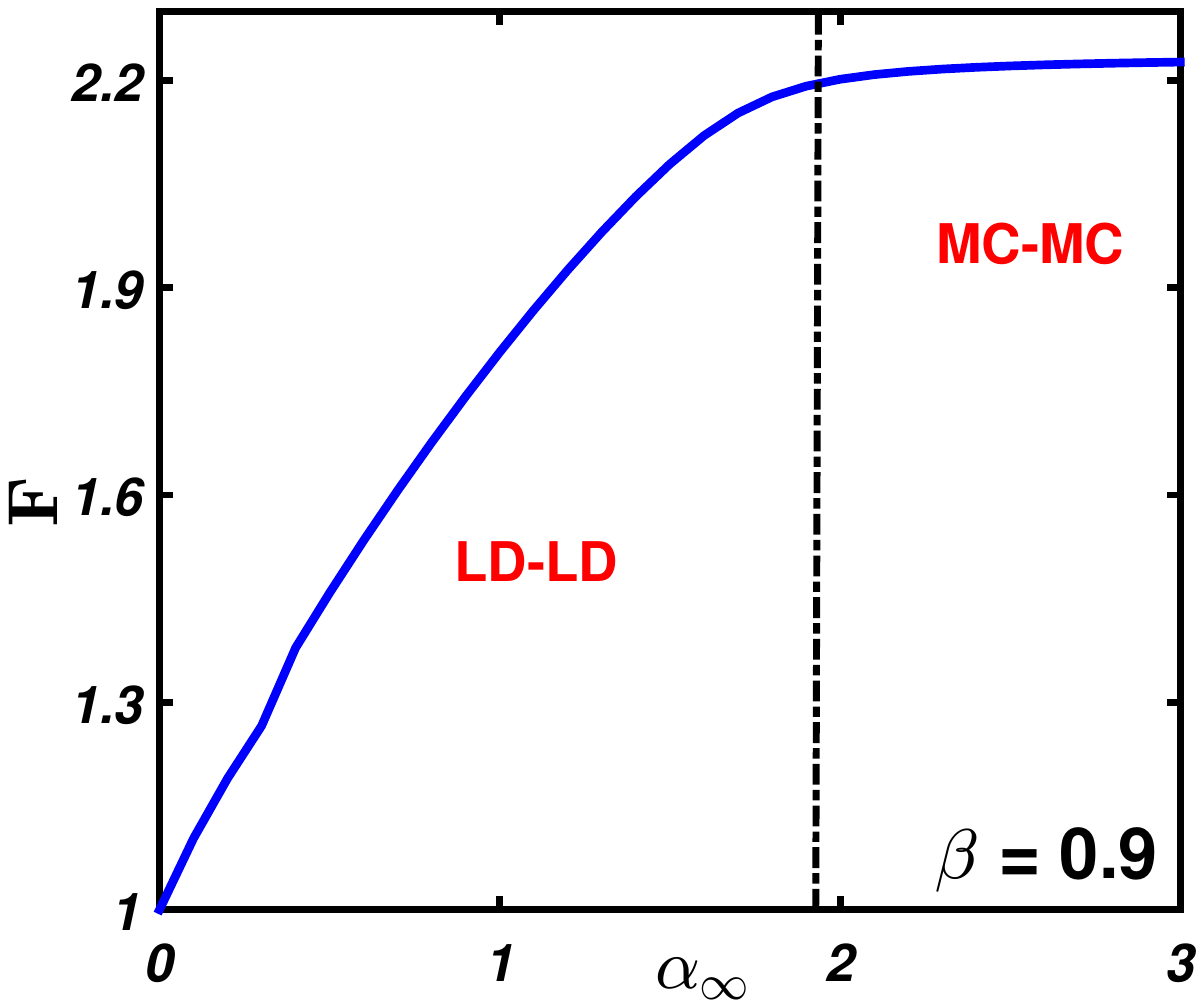}}
 \hspace{0cm} \subfigure[\label{fig7a}]{\includegraphics[trim = 180 230 95 240,width=.28\textwidth,height=5cm]{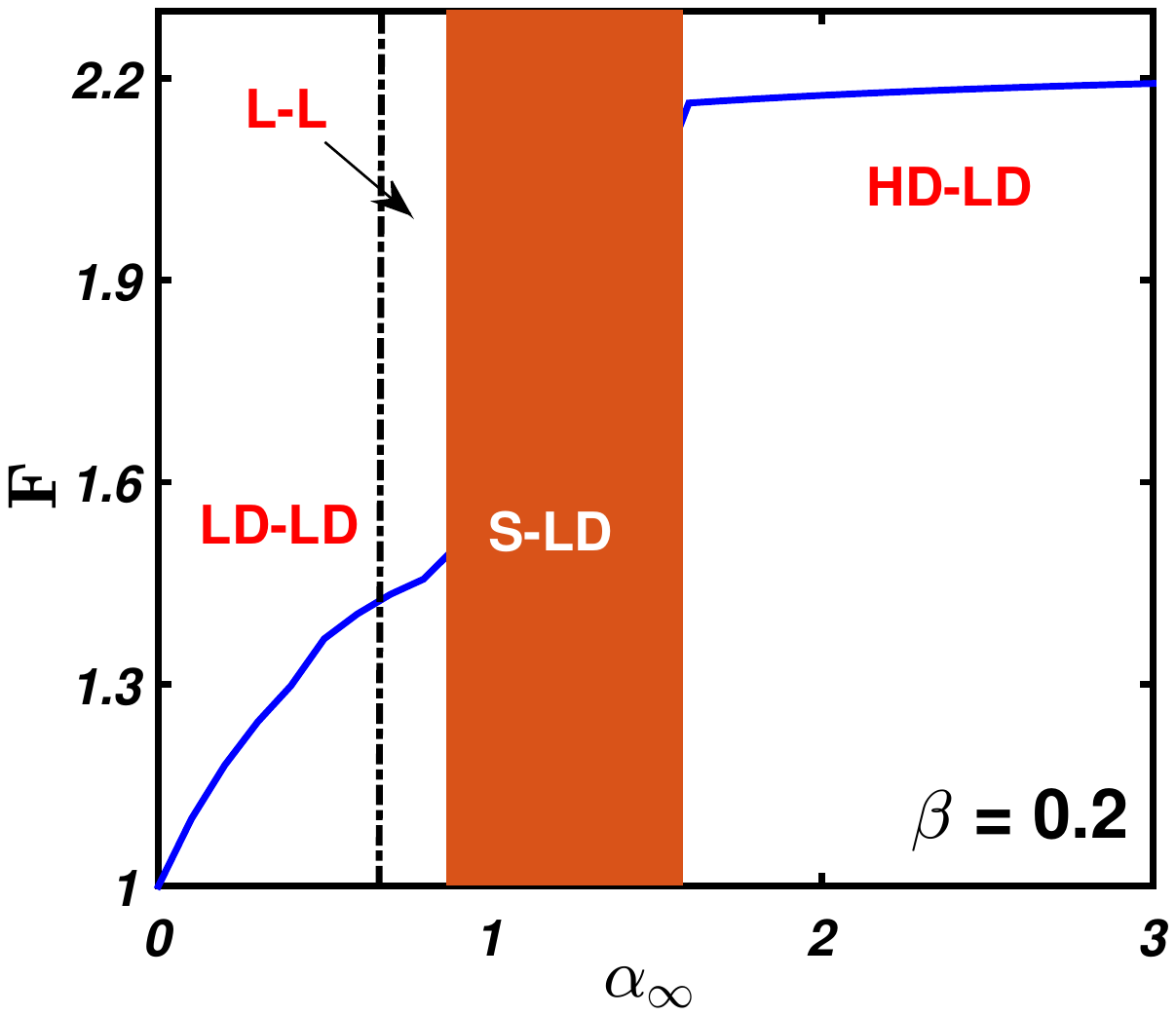}}
            \caption{\label{fig7} Flattening parameter for varied $\alpha_{\infty}$ in different phases with $\Gamma=-2.5$ and $N=25$ in (a) $\beta=0.5$ (b) $\beta=0.9$ (c) $\beta=0.2$.}
\end{figure*}
\begin{figure*}[ht!]
\centering
 \hspace{0cm} \subfigure[]{\includegraphics[trim = 180 230 90 240,width=.28\textwidth,height=5cm]{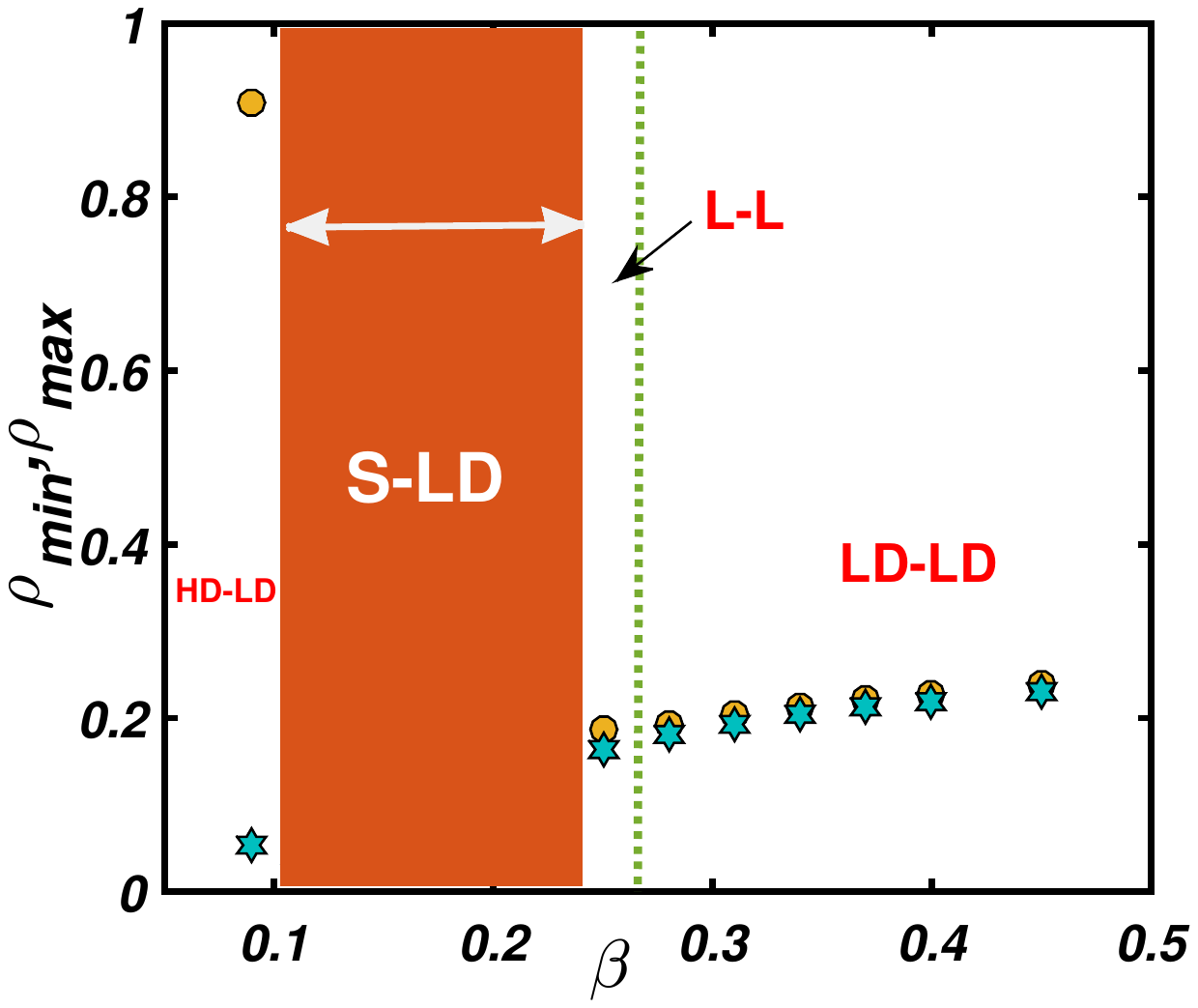}}
  \hspace*{0cm} \subfigure[]{\includegraphics[trim = 180 230 90 240,width=.28\textwidth,height=5cm]{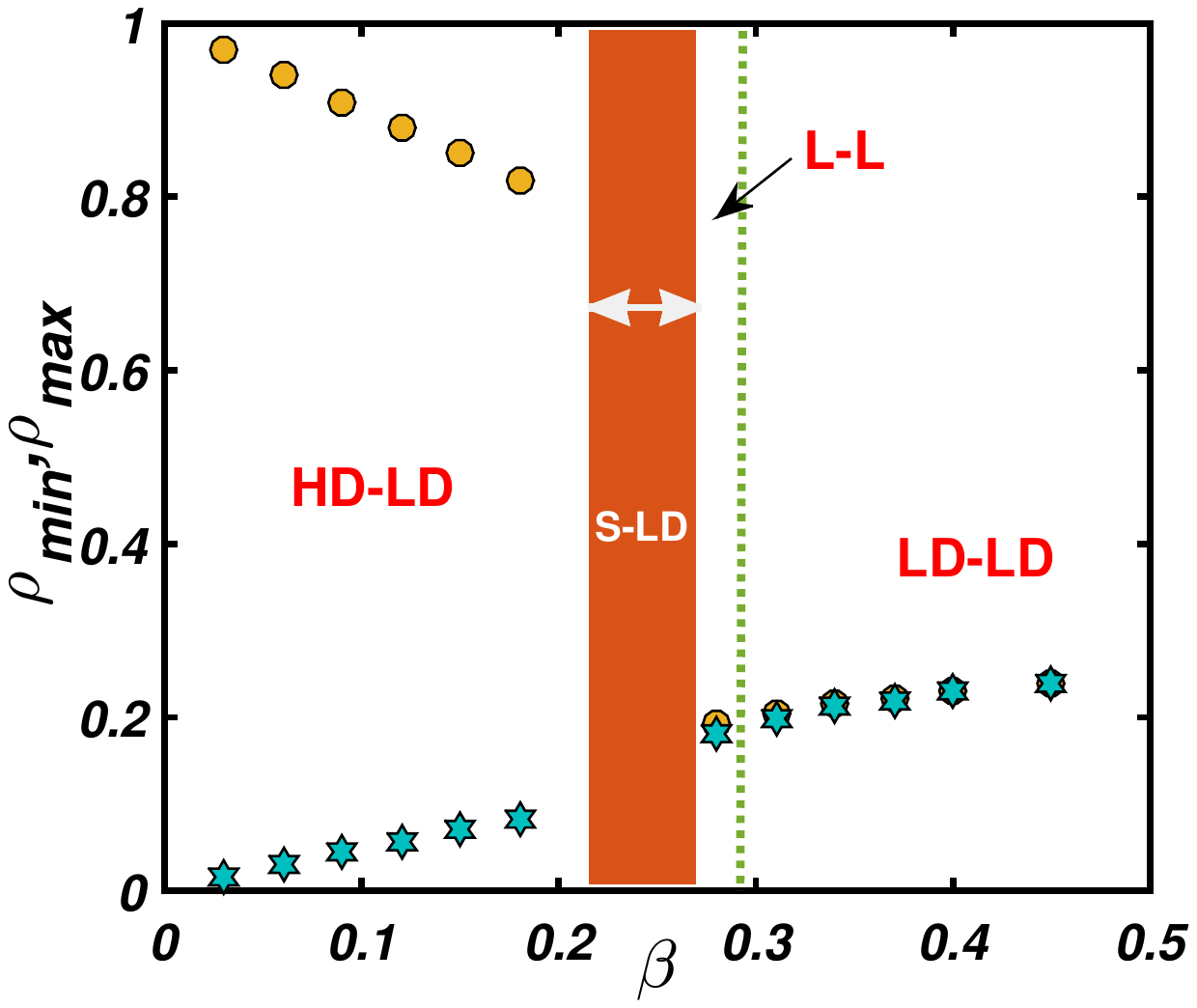}}
  \hspace*{0cm} \subfigure[]{\includegraphics[trim = 180 230 100 240,width=.28\textwidth,height=5cm]{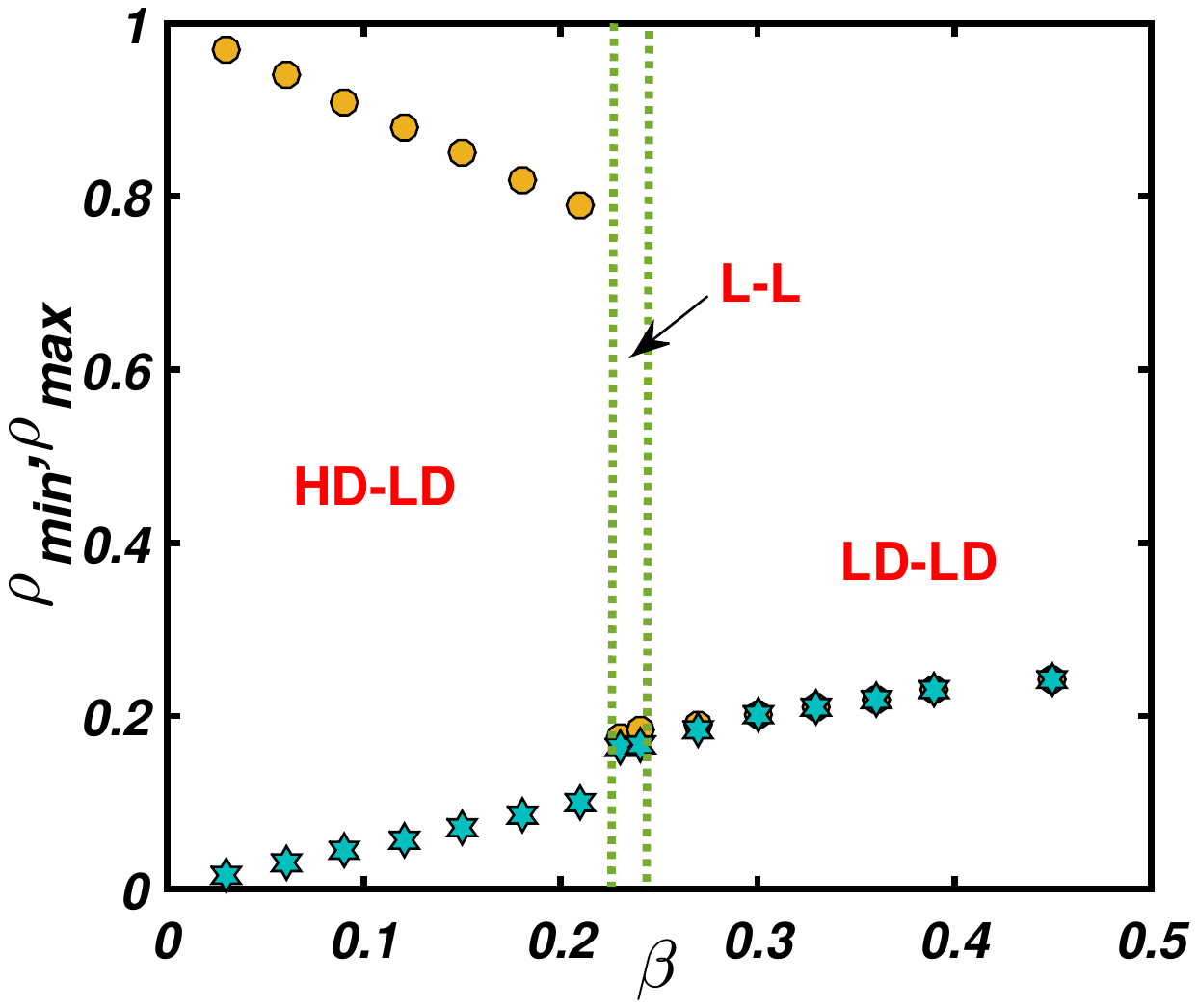}}
            \caption{\label{fig8} Size scaling effect on S-LD phase with increasing (a) $N=25$, (b) $N=100$, (c) $N=800$ for fixed $\alpha_{\infty}=1$ and $\Gamma=-2.5$ with other parameters $\ell=1$, $l_p=0.1$, and $a=1.5$. This figure signifies that with increase in lattice length $N$, S-LD phase shrinks and vanishes in the thermodynamic limit.}
\end{figure*}

With the aim to investigate the role of particle density on the dynamics of system we construct stationary phase diagram for $\Gamma=-2.5$ and $\Gamma=-5$ as shown in Fig.\ref{fig4}. It is evident from the phase diagram that this dynamics produced non-trivial effect on the phase schema. One notices that beyond a critical value $\Gamma\approx-0.4$, a new asymmetrical phase shock-low density (S-LD) phase sandwiched between L-L and HD-LD phase emerges in the system with pre-existing symmetrical phases being intact. Furthermore, decrease in $\Gamma$ spreads LD-LD phase leading to shrinkage in MC-MC and the region where three asymmetric phases exist. Moreover, within the asymmetric regime S-LD phase widens that narrows the HD-LD and L-L phase.

The average density profiles of all the possible steady-state phases with $\Gamma=-5$ are portrayed in Fig.\ref{fig5}. To summarize, the observed phase transitions of a particular type of particle in HD $\rightarrow$ S $\rightarrow$ LD phase with another species being in LD phase are shown in Fig.\ref{fig6}. To visualize the effect of density of particles on flattening parameter $F$ we plotted Fig.\ref{fig7} that illustrates how end-to-end distance varies in different regimes for a fixed value of $\beta$. For $\beta=0.5$ as $\alpha_{\infty}$ increases the flattening parameter of lattice increases linearly because the density of  particles increases. Whereas for $\beta=0.9$, $F$ increases linearly for those value of $\alpha_{\infty}$ that manifests LD-LD region and becomes independent in MC-MC region for which the density of particles is fixed (see Fig.\ref{fig7c}). It is interesting to spot that for $\beta=0.2$, when shock phase persists it is not possible to depict the actual trend of $F$ because the particles are not uniformly distributed on the lattice, therefore, this region is shaded in Fig.\ref{fig7a}. Further increase in $\alpha_{\infty}$ exhibits HD-LD phase where the density of particles is maximum on the lattice that flattens the lattice keeping two ends of the lattice far away from each other. Whereas, in LD-LD phase the lattice is in a compact shape because most of the sites are empty.

\subsection{Size-scaling dependency on asymmetric S-LD phase}
Interestingly, we observe very robust size scaling effects on the newly appeared S-LD phase for the proposed system in the thermodynamic limit. To analyse this effect we denote maximum and minimum of steady-state density of positive and negative particles as $\rho_{max}$ and $\rho_{min}$ respectively. To study this reliance, we probe $\rho_{max}$ and $\rho_{min}$ versus $\beta$ for $\alpha_{\infty}=1$ and $\Gamma=-2.5$ with varying lattice length $N=25$, $N=100$ and $N=800$ as shown in Fig.(\ref{fig8}). As evident from the figure, S-LD phase represented by shaded region shrinks with increase in $N=25$ to $N=100$ followed by expansion in HD-LD phase. Further, in the thermodynamic limit this phase vanishes and HD-LD phases spreads over the S-LD phase. This noticeable feature is well suited to the valid physical arguments. Since, end-to-end distance is directly proportional to $N$, increase in $N$ leads to an increase in $R$. This implies majority of sites are occupied by the particles due to which the lattice is more flattened suggesting the existence of HD-LD phase. Also, it is important to note that for large system size the two spheres at the two ends are far away from each other. In this limit, the system behaves like a bidirectional TASEP with rigid lattice where four phases two symmetric and two asymmetric are observed excluding S-LD phase \cite{evans1995spontaneous,evans1995asymmetric}.
\section{Conclusion}
The transportation of motor proteins of kinesin and dynein family on a flexible polymer like microtubule in a 3D environment provides the motivation for present the study.
For mathematical description, a microtubule is considered to be a lattice of finite length with two distinct types of particles hopping in opposite directions. On the whole, we study a bidirectional totally asymmetric simple exclusion process (TASEP) model where particles hop on a one dimensional flexible lattice interacting only at the boundaries. Additionally, the system is supposed to be immersed in a 3D pool of particles that creates a spherical neighborhood around the entry site influencing the entry rate of particles. Various crucial steady state properties including phase diagrams, density profiles, end-to-end distance of lattice, size scaling effects are thoroughly investigated in the framework of mean-field theory. We even provide the theoretical expressions for the existence of distinct phases in the phase schema.

We scrutinized two different scenarios: (a) The flexible lattice with invariant end-to-end distance and (b) The interaction of moving particles with the conformation of lattice that directs density dependent end-to-end distance. As a consequence, the recycling of particles to the lattice is greatly influenced measured by the parameter called recycling strength. In the former case we observed four stationary phases, two symmetric; low density-low density (LD-LD), maximal current-maximal current (MC-MC) and two asymmetric; low density-low density (L-L), high density-low density (HD-LD) phases. In particular, with varying values of recycling strength the phase boundaries shift in comparison to purely one dimensional bidirectional TASEP. Interestingly for the latter dynamics, a non-trivial effect on the characteristics of system has been reported. Due to an interplay between two distinct species of particles within the constrained entry rate due to local concentration of particles around the entry site, particles display a jam like situation. Hence, in addition to the existing phases, a new asymmetric regime shock-low density (S-LD) emerges in the system. For the better analysis of this additional phase we exploit the phase transitions of a particular species from HD$\rightarrow$S$\rightarrow$LD phase. Furthermore, we exploit the variation of end-to-end distance with the total number particles present on the lattice. Since, in HD-LD phase majority of sites are occupied with particles the lattice is flattened and the two end points of lattice are far away from each other. Whereas, for LD-LD phase the lattice is in its compact shape because most of the sites are empty. In the direction to obtain deeper insight to the existence of S-LD phase, we analyse the system in thermodynamic limit. For large system size this phase diminishes and is dominated by HD-LD phase which is physically relevant. Since, the lattice length and end-to-end distance holds a directly proportional relationship it favors densely populated particles on the lattice.

The proposed work is an attempt to understand the transportation of motor proteins on flexible polymer like microtubule in a 3D environment by highlighting non-trivial effect on the system dynamics. Further, we would like to study a generalized model including Langmuir Kinetics that might introduce additional interesting features to the stationary properties of the system.
\appendix
\section{}\label{appendix}
The continuum part of governing equations is discretized using finite-difference scheme where time and space derivative are replaced using forward and central difference formula. Choosing $\Delta x=\frac{1}{N}$ and suitable $\Delta t$ satisfying the stability criteria $\frac{\Delta t}{\Delta x}\leq 1$, the solution is captured in the limit $j\rightarrow \infty$ (time variable) for $1<i<N$ (space variable) to ensure the occurrence of steady-state.
\begin{eqnarray}
\rho_i^{(j+1)}=\rho_i^{(j)}&&+\dfrac{\epsilon}{2}\dfrac{\Delta t}{\Delta x^2}(\rho_{i+1}^{(j)}-2\rho_i^{(j)}+\rho_{i-1}^{(j)})\\&&\nonumber+\dfrac{\Delta t}{2\Delta x}(2\rho_i^{(j-1)}-1)(\rho_{i+1}^{(j)}-\rho_{i-1}^{(j)},\\
\sigma_i^{(j+1)}=\sigma_i^{(j)}&&+\dfrac{\epsilon}{2}\dfrac{\Delta t}{\Delta x^2}(\sigma_{i+1}^{(j)}-2\rho_i^n+\sigma_{i-1}^{(j)})\\&&\nonumber+\dfrac{\Delta t}{2\Delta x}(1-2\sigma_i^{(j)})(\sigma_{i+1}^{(j)}-\sigma_{i-1}^{(j)})
\end{eqnarray}

\begin{thebibliography}{28}%
\makeatletter
\providecommand \@ifxundefined [1]{%
 \@ifx{#1\undefined}
}%
\providecommand \@ifnum [1]{%
 \ifnum #1\expandafter \@firstoftwo
 \else \expandafter \@secondoftwo
 \fi
}%
\providecommand \@ifx [1]{%
 \ifx #1\expandafter \@firstoftwo
 \else \expandafter \@secondoftwo
 \fi
}%
\providecommand \natexlab [1]{#1}%
\providecommand \enquote  [1]{``#1''}%
\providecommand \bibnamefont  [1]{#1}%
\providecommand \bibfnamefont [1]{#1}%
\providecommand \citenamefont [1]{#1}%
\providecommand \href@noop [0]{\@secondoftwo}%
\providecommand \href [0]{\begingroup \@sanitize@url \@href}%
\providecommand \@href[1]{\@@startlink{#1}\@@href}%
\providecommand \@@href[1]{\endgroup#1\@@endlink}%
\providecommand \@sanitize@url [0]{\catcode `\\12\catcode `\$12\catcode
  `\&12\catcode `\#12\catcode `\^12\catcode `\_12\catcode `\%12\relax}%
\providecommand \@@startlink[1]{}%
\providecommand \@@endlink[0]{}%
\providecommand \url  [0]{\begingroup\@sanitize@url \@url }%
\providecommand \@url [1]{\endgroup\@href {#1}{\urlprefix }}%
\providecommand \urlprefix  [0]{URL }%
\providecommand \Eprint [0]{\href }%
\providecommand \doibase [0]{http://dx.doi.org/}%
\providecommand \selectlanguage [0]{\@gobble}%
\providecommand \bibinfo  [0]{\@secondoftwo}%
\providecommand \bibfield  [0]{\@secondoftwo}%
\providecommand \translation [1]{[#1]}%
\providecommand \BibitemOpen [0]{}%
\providecommand \bibitemStop [0]{}%
\providecommand \bibitemNoStop [0]{.\EOS\space}%
\providecommand \EOS [0]{\spacefactor3000\relax}%
\providecommand \BibitemShut  [1]{\csname bibitem#1\endcsname}%
\let\auto@bib@innerbib\@empty
\bibitem [{\citenamefont {Widom}\ \emph {et~al.}(1991)\citenamefont {Widom},
  \citenamefont {Viovy},\ and\ \citenamefont {Defontaines}}]{widom1991repton}%
  \BibitemOpen
  \bibfield  {author} {\bibinfo {author} {\bibfnamefont {B.}~\bibnamefont
  {Widom}}, \bibinfo {author} {\bibfnamefont {J.}~\bibnamefont {Viovy}}, \ and\
  \bibinfo {author} {\bibfnamefont {A.}~\bibnamefont {Defontaines}},\
  }\href@noop {} {\bibfield  {journal} {\bibinfo  {journal} {Journal de
  Physique I}\ }\textbf {\bibinfo {volume} {1}},\ \bibinfo {pages} {1759}
  (\bibinfo {year} {1991})}\BibitemShut {NoStop}%
\bibitem [{\citenamefont {Chowdhury}\ \emph {et~al.}(2000)\citenamefont
  {Chowdhury}, \citenamefont {Santen},\ and\ \citenamefont
  {Schadschneider}}]{chowdhury2000statistical}%
  \BibitemOpen
  \bibfield  {author} {\bibinfo {author} {\bibfnamefont {D.}~\bibnamefont
  {Chowdhury}}, \bibinfo {author} {\bibfnamefont {L.}~\bibnamefont {Santen}}, \
  and\ \bibinfo {author} {\bibfnamefont {A.}~\bibnamefont {Schadschneider}},\
  }\href@noop {} {\bibfield  {journal} {\bibinfo  {journal} {Physics Reports}\
  }\textbf {\bibinfo {volume} {329}},\ \bibinfo {pages} {199} (\bibinfo {year}
  {2000})}\BibitemShut {NoStop}%
\bibitem [{\citenamefont {Belitsky}\ \emph {et~al.}(2001)\citenamefont
  {Belitsky}, \citenamefont {Krug}, \citenamefont {Neves},\ and\ \citenamefont
  {Sch{\"u}tz}}]{belitsky2001cellular}%
  \BibitemOpen
  \bibfield  {author} {\bibinfo {author} {\bibfnamefont {V.}~\bibnamefont
  {Belitsky}}, \bibinfo {author} {\bibfnamefont {J.}~\bibnamefont {Krug}},
  \bibinfo {author} {\bibfnamefont {E.~J.}\ \bibnamefont {Neves}}, \ and\
  \bibinfo {author} {\bibfnamefont {G.}~\bibnamefont {Sch{\"u}tz}},\
  }\href@noop {} {\bibfield  {journal} {\bibinfo  {journal} {Journal of
  Statistical Physics}\ }\textbf {\bibinfo {volume} {103}},\ \bibinfo {pages}
  {945} (\bibinfo {year} {2001})}\BibitemShut {NoStop}%
\bibitem [{\citenamefont {Foulaadvand}\ and\ \citenamefont
  {Maass}(2016)}]{foulaadvand2016phase}%
  \BibitemOpen
  \bibfield  {author} {\bibinfo {author} {\bibfnamefont {M.~E.}\ \bibnamefont
  {Foulaadvand}}\ and\ \bibinfo {author} {\bibfnamefont {P.}~\bibnamefont
  {Maass}},\ }\href@noop {} {\bibfield  {journal} {\bibinfo  {journal}
  {Physical Review E}\ }\textbf {\bibinfo {volume} {94}},\ \bibinfo {pages}
  {012304} (\bibinfo {year} {2016})}\BibitemShut {NoStop}%
\bibitem [{\citenamefont {MacDonald}\ \emph {et~al.}(1968)\citenamefont
  {MacDonald}, \citenamefont {Gibbs},\ and\ \citenamefont
  {Pipkin}}]{macdonald1968kinetics}%
  \BibitemOpen
  \bibfield  {author} {\bibinfo {author} {\bibfnamefont {C.~T.}\ \bibnamefont
  {MacDonald}}, \bibinfo {author} {\bibfnamefont {J.~H.}\ \bibnamefont
  {Gibbs}}, \ and\ \bibinfo {author} {\bibfnamefont {A.~C.}\ \bibnamefont
  {Pipkin}},\ }\href@noop {} {\bibfield  {journal} {\bibinfo  {journal}
  {Biopolymers: Original Research on Biomolecules}\ }\textbf {\bibinfo {volume}
  {6}},\ \bibinfo {pages} {1} (\bibinfo {year} {1968})}\BibitemShut {NoStop}%
\bibitem [{\citenamefont {Blythe}\ and\ \citenamefont
  {Evans}(2007)}]{blythe2007nonequilibrium}%
  \BibitemOpen
  \bibfield  {author} {\bibinfo {author} {\bibfnamefont {R.~A.}\ \bibnamefont
  {Blythe}}\ and\ \bibinfo {author} {\bibfnamefont {M.~R.}\ \bibnamefont
  {Evans}},\ }\href@noop {} {\bibfield  {journal} {\bibinfo  {journal} {Journal
  of Physics A: Mathematical and Theoretical}\ }\textbf {\bibinfo {volume}
  {40}},\ \bibinfo {pages} {R333} (\bibinfo {year} {2007})}\BibitemShut
  {NoStop}%
\bibitem [{\citenamefont {Zia}\ \emph {et~al.}(2011)\citenamefont {Zia},
  \citenamefont {Dong},\ and\ \citenamefont {Schmittmann}}]{zia2011modeling}%
  \BibitemOpen
  \bibfield  {author} {\bibinfo {author} {\bibfnamefont {R.}~\bibnamefont
  {Zia}}, \bibinfo {author} {\bibfnamefont {J.}~\bibnamefont {Dong}}, \ and\
  \bibinfo {author} {\bibfnamefont {B.}~\bibnamefont {Schmittmann}},\
  }\href@noop {} {\bibfield  {journal} {\bibinfo  {journal} {Journal of
  Statistical Physics}\ }\textbf {\bibinfo {volume} {144}},\ \bibinfo {pages}
  {405} (\bibinfo {year} {2011})}\BibitemShut {NoStop}%
\bibitem [{\citenamefont {Kolomeisky}\ \emph {et~al.}(1998)\citenamefont
  {Kolomeisky}, \citenamefont {Sch{\"u}tz}, \citenamefont {Kolomeisky},\ and\
  \citenamefont {Straley}}]{kolomeisky1998phase}%
  \BibitemOpen
  \bibfield  {author} {\bibinfo {author} {\bibfnamefont {A.~B.}\ \bibnamefont
  {Kolomeisky}}, \bibinfo {author} {\bibfnamefont {G.~M.}\ \bibnamefont
  {Sch{\"u}tz}}, \bibinfo {author} {\bibfnamefont {E.~B.}\ \bibnamefont
  {Kolomeisky}}, \ and\ \bibinfo {author} {\bibfnamefont {J.~P.}\ \bibnamefont
  {Straley}},\ }\href@noop {} {\bibfield  {journal} {\bibinfo  {journal}
  {Journal of Physics A: Mathematical and General}\ }\textbf {\bibinfo {volume}
  {31}},\ \bibinfo {pages} {6911} (\bibinfo {year} {1998})}\BibitemShut
  {NoStop}%
\bibitem [{\citenamefont {Evans}\ \emph {et~al.}(2003)\citenamefont {Evans},
  \citenamefont {Juh{\'a}sz},\ and\ \citenamefont {Santen}}]{evans2003shock}%
  \BibitemOpen
  \bibfield  {author} {\bibinfo {author} {\bibfnamefont {M.~R.}\ \bibnamefont
  {Evans}}, \bibinfo {author} {\bibfnamefont {R.}~\bibnamefont {Juh{\'a}sz}}, \
  and\ \bibinfo {author} {\bibfnamefont {L.}~\bibnamefont {Santen}},\
  }\href@noop {} {\bibfield  {journal} {\bibinfo  {journal} {Physical Review
  E}\ }\textbf {\bibinfo {volume} {68}},\ \bibinfo {pages} {026117} (\bibinfo
  {year} {2003})}\BibitemShut {NoStop}%
\bibitem [{\citenamefont {Soppina}\ \emph {et~al.}(2009)\citenamefont
  {Soppina}, \citenamefont {Rai}, \citenamefont {Ramaiya}, \citenamefont
  {Barak},\ and\ \citenamefont {Mallik}}]{soppina2009tug}%
  \BibitemOpen
  \bibfield  {author} {\bibinfo {author} {\bibfnamefont {V.}~\bibnamefont
  {Soppina}}, \bibinfo {author} {\bibfnamefont {A.~K.}\ \bibnamefont {Rai}},
  \bibinfo {author} {\bibfnamefont {A.~J.}\ \bibnamefont {Ramaiya}}, \bibinfo
  {author} {\bibfnamefont {P.}~\bibnamefont {Barak}}, \ and\ \bibinfo {author}
  {\bibfnamefont {R.}~\bibnamefont {Mallik}},\ }\href@noop {} {\bibfield
  {journal} {\bibinfo  {journal} {Proceedings of the National Academy of
  Sciences}\ }\textbf {\bibinfo {volume} {106}},\ \bibinfo {pages} {19381}
  (\bibinfo {year} {2009})}\BibitemShut {NoStop}%
\bibitem [{\citenamefont {Hancock}(2014)}]{hancock2014bidirectional}%
  \BibitemOpen
  \bibfield  {author} {\bibinfo {author} {\bibfnamefont {W.~O.}\ \bibnamefont
  {Hancock}},\ }\href@noop {} {\bibfield  {journal} {\bibinfo  {journal}
  {Nature reviews Molecular cell biology}\ }\textbf {\bibinfo {volume} {15}},\
  \bibinfo {pages} {615} (\bibinfo {year} {2014})}\BibitemShut {NoStop}%
\bibitem [{\citenamefont {Schliwa}\ and\ \citenamefont
  {Woehlke}(2003)}]{schliwa2003molecular}%
  \BibitemOpen
  \bibfield  {author} {\bibinfo {author} {\bibfnamefont {M.}~\bibnamefont
  {Schliwa}}\ and\ \bibinfo {author} {\bibfnamefont {G.}~\bibnamefont
  {Woehlke}},\ }\href@noop {} {\bibfield  {journal} {\bibinfo  {journal}
  {Nature}\ }\textbf {\bibinfo {volume} {422}},\ \bibinfo {pages} {759}
  (\bibinfo {year} {2003})}\BibitemShut {NoStop}%
\bibitem [{\citenamefont {Evans}\ \emph
  {et~al.}(1995{\natexlab{a}})\citenamefont {Evans}, \citenamefont {Foster},
  \citenamefont {Godr{\`e}che},\ and\ \citenamefont
  {Mukamel}}]{evans1995spontaneous}%
  \BibitemOpen
  \bibfield  {author} {\bibinfo {author} {\bibfnamefont {M.~R.}\ \bibnamefont
  {Evans}}, \bibinfo {author} {\bibfnamefont {D.~P.}\ \bibnamefont {Foster}},
  \bibinfo {author} {\bibfnamefont {C.}~\bibnamefont {Godr{\`e}che}}, \ and\
  \bibinfo {author} {\bibfnamefont {D.}~\bibnamefont {Mukamel}},\ }\href@noop
  {} {\bibfield  {journal} {\bibinfo  {journal} {Physical review letters}\
  }\textbf {\bibinfo {volume} {74}},\ \bibinfo {pages} {208} (\bibinfo {year}
  {1995}{\natexlab{a}})}\BibitemShut {NoStop}%
\bibitem [{\citenamefont {Sharma}\ and\ \citenamefont
  {Gupta}(2017)}]{sharma2017phase}%
  \BibitemOpen
  \bibfield  {author} {\bibinfo {author} {\bibfnamefont {N.}~\bibnamefont
  {Sharma}}\ and\ \bibinfo {author} {\bibfnamefont {A.}~\bibnamefont {Gupta}},\
  }\href@noop {} {\bibfield  {journal} {\bibinfo  {journal} {Journal of
  Statistical Mechanics: Theory and Experiment}\ }\textbf {\bibinfo {volume}
  {2017}},\ \bibinfo {pages} {043211} (\bibinfo {year} {2017})}\BibitemShut
  {NoStop}%
\bibitem [{\citenamefont {Verma}\ \emph {et~al.}(2018)\citenamefont {Verma},
  \citenamefont {Sharma},\ and\ \citenamefont {Gupta}}]{verma2018far}%
  \BibitemOpen
  \bibfield  {author} {\bibinfo {author} {\bibfnamefont {A.~K.}\ \bibnamefont
  {Verma}}, \bibinfo {author} {\bibfnamefont {N.}~\bibnamefont {Sharma}}, \
  and\ \bibinfo {author} {\bibfnamefont {A.~K.}\ \bibnamefont {Gupta}},\
  }\href@noop {} {\bibfield  {journal} {\bibinfo  {journal} {Physical Review
  E}\ }\textbf {\bibinfo {volume} {97}},\ \bibinfo {pages} {022105} (\bibinfo
  {year} {2018})}\BibitemShut {NoStop}%
\bibitem [{\citenamefont {Evans}\ \emph
  {et~al.}(1995{\natexlab{b}})\citenamefont {Evans}, \citenamefont {Foster},
  \citenamefont {Godreche},\ and\ \citenamefont
  {Mukamel}}]{evans1995asymmetric}%
  \BibitemOpen
  \bibfield  {author} {\bibinfo {author} {\bibfnamefont {M.}~\bibnamefont
  {Evans}}, \bibinfo {author} {\bibfnamefont {D.}~\bibnamefont {Foster}},
  \bibinfo {author} {\bibfnamefont {C.}~\bibnamefont {Godreche}}, \ and\
  \bibinfo {author} {\bibfnamefont {D.}~\bibnamefont {Mukamel}},\ }\href@noop
  {} {\bibfield  {journal} {\bibinfo  {journal} {Journal of statistical
  physics}\ }\textbf {\bibinfo {volume} {80}},\ \bibinfo {pages} {69} (\bibinfo
  {year} {1995}{\natexlab{b}})}\BibitemShut {NoStop}%
\bibitem [{\citenamefont {Alberts}\ \emph {et~al.}(2008)\citenamefont
  {Alberts}, \citenamefont {Johnson}, \citenamefont {Lewis}, \citenamefont
  {Raff}, \citenamefont {Roberts},\ and\ \citenamefont
  {Walter}}]{alberts2008molecular}%
  \BibitemOpen
  \bibfield  {author} {\bibinfo {author} {\bibfnamefont {B.}~\bibnamefont
  {Alberts}}, \bibinfo {author} {\bibfnamefont {A.}~\bibnamefont {Johnson}},
  \bibinfo {author} {\bibfnamefont {J.}~\bibnamefont {Lewis}}, \bibinfo
  {author} {\bibfnamefont {M.}~\bibnamefont {Raff}}, \bibinfo {author}
  {\bibfnamefont {K.}~\bibnamefont {Roberts}}, \ and\ \bibinfo {author}
  {\bibfnamefont {P.}~\bibnamefont {Walter}},\ }\href@noop {} {\bibfield
  {journal} {\bibinfo  {journal} {New York}\ } (\bibinfo {year}
  {2008})}\BibitemShut {NoStop}%
\bibitem [{\citenamefont {Gosselin}\ \emph {et~al.}(2016)\citenamefont
  {Gosselin}, \citenamefont {Mohrbach}, \citenamefont {Kuli{\'c}},\ and\
  \citenamefont {Ziebert}}]{gosselin2016complex}%
  \BibitemOpen
  \bibfield  {author} {\bibinfo {author} {\bibfnamefont {P.}~\bibnamefont
  {Gosselin}}, \bibinfo {author} {\bibfnamefont {H.}~\bibnamefont {Mohrbach}},
  \bibinfo {author} {\bibfnamefont {I.~M.}\ \bibnamefont {Kuli{\'c}}}, \ and\
  \bibinfo {author} {\bibfnamefont {F.}~\bibnamefont {Ziebert}},\ }\href@noop
  {} {\bibfield  {journal} {\bibinfo  {journal} {Physica D: Nonlinear
  Phenomena}\ }\textbf {\bibinfo {volume} {318}},\ \bibinfo {pages} {105}
  (\bibinfo {year} {2016})}\BibitemShut {NoStop}%
\bibitem [{\citenamefont {Howard}\ \emph {et~al.}(2001)\citenamefont {Howard}
  \emph {et~al.}}]{howard2001mechanics}%
  \BibitemOpen
   \bibfield  {author} {\bibinfo {author} {\bibfnamefont {J.}~\bibnamefont
  {Howard}} \emph {et~al.},\ }\href@noop {} {\bibfield  {journal} {\bibinfo  {journal} {Mechanics of motor proteins and the cytoskeleton (Sinauer associates Sunderland, MA)}\ }\  (\bibinfo {year}
  {2001})}\BibitemShut {NoStop}%
\bibitem [{\citenamefont {Cai}\ \emph {et~al.}(2001)\citenamefont {Cai},
  \citenamefont {Romagnoli},\ and\ \citenamefont
  {Cresti}}]{cai2001microtubule}%
  \BibitemOpen
  \bibfield  {author} {\bibinfo {author} {\bibfnamefont {G.}~\bibnamefont
  {Cai}}, \bibinfo {author} {\bibfnamefont {S.}~\bibnamefont {Romagnoli}}, \
  and\ \bibinfo {author} {\bibfnamefont {M.}~\bibnamefont {Cresti}},\
  }\href@noop {} {\bibfield  {journal} {\bibinfo  {journal} {Sexual Plant
  Reproduction}\ }\textbf {\bibinfo {volume} {14}},\ \bibinfo {pages} {27}
  (\bibinfo {year} {2001})}\BibitemShut {NoStop}%
\bibitem [{\citenamefont {Luby-Phelps}\ \emph {et~al.}(1986)\citenamefont
  {Luby-Phelps}, \citenamefont {Taylor},\ and\ \citenamefont
  {Lanni}}]{luby1986probing}%
  \BibitemOpen
  \bibfield  {author} {\bibinfo {author} {\bibfnamefont {K.}~\bibnamefont
  {Luby-Phelps}}, \bibinfo {author} {\bibfnamefont {D.~L.}\ \bibnamefont
  {Taylor}}, \ and\ \bibinfo {author} {\bibfnamefont {F.}~\bibnamefont
  {Lanni}},\ }\href@noop {} {\bibfield  {journal} {\bibinfo  {journal} {The
  Journal of cell biology}\ }\textbf {\bibinfo {volume} {102}},\ \bibinfo
  {pages} {2015} (\bibinfo {year} {1986})}\BibitemShut {NoStop}%
\bibitem [{\citenamefont {Fernandes}\ and\ \citenamefont
  {Ciandrini}(2019)}]{fernandes2019driven}%
  \BibitemOpen
  \bibfield  {author} {\bibinfo {author} {\bibfnamefont {L.~D.}\ \bibnamefont
  {Fernandes}}\ and\ \bibinfo {author} {\bibfnamefont {L.}~\bibnamefont
  {Ciandrini}},\ }\href@noop {} {\bibfield  {journal} {\bibinfo  {journal}
  {Physical Review E}\ }\textbf {\bibinfo {volume} {99}},\ \bibinfo {pages}
  {052409} (\bibinfo {year} {2019})}\BibitemShut {NoStop}%
\bibitem [{\citenamefont {Fernandes}\ \emph {et~al.}(2017)\citenamefont
  {Fernandes}, \citenamefont {De~Moura},\ and\ \citenamefont
  {Ciandrini}}]{fernandes2017gene}%
  \BibitemOpen
  \bibfield  {author} {\bibinfo {author} {\bibfnamefont {L.~D.}\ \bibnamefont
  {Fernandes}}, \bibinfo {author} {\bibfnamefont {A.~P.}\ \bibnamefont
  {De~Moura}}, \ and\ \bibinfo {author} {\bibfnamefont {L.}~\bibnamefont
  {Ciandrini}},\ }\href@noop {} {\bibfield  {journal} {\bibinfo  {journal}
  {Scientific reports}\ }\textbf {\bibinfo {volume} {7}},\ \bibinfo {pages} {1}
  (\bibinfo {year} {2017})}\BibitemShut {NoStop}%
\bibitem [{\citenamefont {Verma}\ and\ \citenamefont
  {Gupta}(2019)}]{verma2019stochastic}%
  \BibitemOpen
  \bibfield  {author} {\bibinfo {author} {\bibfnamefont {A.~K.}\ \bibnamefont
  {Verma}}\ and\ \bibinfo {author} {\bibfnamefont {A.~K.}\ \bibnamefont
  {Gupta}},\ }\href@noop {} {\bibfield  {journal} {\bibinfo  {journal} {Journal
  of Statistical Mechanics: Theory and Experiment}\ }\textbf {\bibinfo {volume}
  {2019}},\ \bibinfo {pages} {103210} (\bibinfo {year} {2019})}\BibitemShut
  {NoStop}%
\bibitem [{\citenamefont {Bakshi}\ \emph {et~al.}(2012)\citenamefont {Bakshi},
  \citenamefont {Siryaporn}, \citenamefont {Goulian},\ and\ \citenamefont
  {Weisshaar}}]{bakshi2012superresolution}%
  \BibitemOpen
  \bibfield  {author} {\bibinfo {author} {\bibfnamefont {S.}~\bibnamefont
  {Bakshi}}, \bibinfo {author} {\bibfnamefont {A.}~\bibnamefont {Siryaporn}},
  \bibinfo {author} {\bibfnamefont {M.}~\bibnamefont {Goulian}}, \ and\
  \bibinfo {author} {\bibfnamefont {J.~C.}\ \bibnamefont {Weisshaar}},\
  }\href@noop {} {\bibfield  {journal} {\bibinfo  {journal} {Molecular
  microbiology}\ }\textbf {\bibinfo {volume} {85}},\ \bibinfo {pages} {21}
  (\bibinfo {year} {2012})}\BibitemShut {NoStop}%
\bibitem [{\citenamefont {Eisler}(1969)}]{eisler1969introduction}%
  \BibitemOpen
  \bibfield  {author} {\bibinfo {author} {\bibfnamefont {T.~J.}\ \bibnamefont
  {Eisler}},\ }\href@noop {} {\emph {\bibinfo {title} {An Introduction to
  Green's Functions}}},\ \bibinfo {type} {Tech. Rep.}\ (\bibinfo  {institution}
  {Catholic Univ Of America Washington Dc Inst Of Ocean Science And
  Engineering},\ \bibinfo {year} {1969})\BibitemShut {NoStop}%
\bibitem [{\citenamefont {Derrida}\ \emph {et~al.}(1992)\citenamefont
  {Derrida}, \citenamefont {Domany},\ and\ \citenamefont
  {Mukamel}}]{derrida1992exact}%
  \BibitemOpen
  \bibfield  {author} {\bibinfo {author} {\bibfnamefont {B.}~\bibnamefont
  {Derrida}}, \bibinfo {author} {\bibfnamefont {E.}~\bibnamefont {Domany}}, \
  and\ \bibinfo {author} {\bibfnamefont {D.}~\bibnamefont {Mukamel}},\
  }\href@noop {} {\bibfield  {journal} {\bibinfo  {journal} {Journal of
  Statistical Physics}\ }\textbf {\bibinfo {volume} {69}},\ \bibinfo {pages}
  {667} (\bibinfo {year} {1992})}\BibitemShut {NoStop}%
\bibitem [{\citenamefont {Gupta}\ and\ \citenamefont
  {Dhiman}(2014)}]{gupta2014asymmetric}%
  \BibitemOpen
  \bibfield  {author} {\bibinfo {author} {\bibfnamefont {A.~K.}\ \bibnamefont
  {Gupta}}\ and\ \bibinfo {author} {\bibfnamefont {I.}~\bibnamefont {Dhiman}},\
  }\href@noop {} {\bibfield  {journal} {\bibinfo  {journal} {Physical Review
  E}\ }\textbf {\bibinfo {volume} {89}},\ \bibinfo {pages} {022131} (\bibinfo
  {year} {2014})}\BibitemShut {NoStop}%
\end{thebibliography}

%

\end{document}